\documentclass[12pt]{article}
\pdfoutput=1

\usepackage{amsmath,amssymb,graphicx,slashed}

\newcommand{\refeq}[1]{Eq.\ (\ref{#1})}

\newcommand{\nn}{\nonumber \\}

\def\beq{\begin{equation}}
\def\eeq{\end{equation}}
\newcommand{\ba}{\begin{array}}
\newcommand{\ea}{\end{array}}
\newcommand{\bea}{\begin{eqnarray}}
\newcommand{\eea}{\end{eqnarray} }
\newcommand{\bal}{\begin{align}}
\newcommand{\eal}{\end{align}}

\newcommand{\Zp}{Z$^\prime$}
\newcommand{\Wp}{W$^\prime$}

\newcommand{\U}[1]{U(#1)}
\newcommand{\SU}[1]{SU(#1)}
\newcommand{\calL}{\mathcal{L}}
\newcommand{\calW}{{\mathcal{W}}}
\newcommand{\calB}{{\mathcal{B}}}

\newcommand{\WpL}{W$^\prime_L$}
\newcommand{\WpR}{W$^\prime_R$}
\newcommand{\ZpL}{Z$^\prime_L$}

\def\MET{\slashed{E}_T}
\def\Tr{\mbox{Tr}}
\def\suma{\displaystyle\sum}
\newcommand{\tbar}{\bar{t}}

\newcommand{\GeV}{~\mathrm{GeV}}
\newcommand{\TeV}{~\mathrm{TeV}}

\newcommand{\tinymath}[1]{{\tiny{\mbox{$#1$}}}}

\def\math@ccstyles#1#2#3#4#5#6#7{{\leavevmode
      \setbox0\mathbox{#6#7}%
      \setbox2\mathbox{#4#5}%
      \dimen@ #3%
      \baselineskip\z@\lineskiplimit#1\lineskip\z@
      \vbox{\ialign{##\crcr
             \hfil \kern #2\box2 \hfil\crcr
             \noalign{\kern\dimen@}%
             \hfil\box0\hfil\crcr}}}}
\def\mathaccstyles{\math@ccstyles\maxdimen}
\def\maththroughstyles{\math@ccstyles{-\maxdimen}}
\def\unity%
 {\maththroughstyles{.45\ht0}\z@\displaystyle {\mathchar"006C}\displaystyle 1}

\newcommand{\half}{\frac{1}{2}}

\begin{document}

\vspace*{2cm}
\begin{center}
{\Large \bf Combining searches of {\Zp} and  {\Wp} bosons}
\vspace*{0.1cm}
\vskip 1cm

\renewcommand{\thefootnote}{\fnsymbol{footnote}}
\setcounter{footnote}{1}

{\large J.\ de Blas$^a$\footnote{jdeblasm@nd.edu}, J. M.\ Lizana$^b$\footnote{jlizan@ugr.es} and M.\ P\'erez-Victoria$^b$\footnote{mpv@ugr.es}}
\vskip 8pt

{$^a$ \it Department of Physics, University of Notre Dame, \\ 
             Notre Dame, IN 46556, USA}\\
{$^b$ \it CAFPE and Departamento de F\'{\i}sica Te\'orica y del Cosmos, \\ 
             Universidad de Granada, E-18071, Spain}\\

\vspace*{0.5cm}

\end{center}


\begin{center}
{\large \bf Abstract}

\vspace{.6cm}

\parbox{14cm}{We study in a model-independent way new neutral and charged vector bosons that could give observable signals with leptonic final states at the LHC. We show, in particular, that a charged vector \Wp{} decaying into lepton plus neutrino is accompanied by at least an extra neutral vector boson \Zp,  nearly degenerate with the charged one. Conversely, a \Zp{} boson with significant isospin violation cannot exist without a companion \Wp{}. To take advantage of these generic correlations, we perform a combined analysis of LHC data in the dilepton and lepton-plus-missing-energy channels, which allows us to improve the limits from independent analyses. We also develop some tools to easily deal with cases in which several heavy vector bosons with similar masses interfere. Finally, we develop a theoretically consistent framework for the study of the sequential \Zp{} and \Wp{} benchmarks.}

\end{center}

\vspace{2cm}

\renewcommand{\thefootnote}{\arabic{footnote}}
\setcounter{footnote}{0}

\newpage


\section{Introduction}
\label{sec:i} 

New vector bosons, i.e.\ new particles of spin 1, are a common occurrence in theories beyond the Standard Model (SM). They appear whenever the gauge group of the SM is extended, as the gauge bosons of the extra broken symmetries. This is the case of Grand Unified Theories, including string constructions, or Little Higgs models. In theories in extra dimensions, the gauge group is often higher dimensional, which gives rise to infinite towers of vector excitations. Finally, vector resonances are a typical feature of theories with a strongly-coupled sector, such as composite Higgs models. The last two scenarios can be related by gauge/gravity duality.

These extra vector bosons, if not too heavy, could give clear signals at hadron colliders. In particular, new neutral and charged vector bosons (\Zp{} and \Wp{}, respectively) with significant couplings to quarks and leptons are most easily seen as resonances in dilepton ($\ell^+ \ell^-$) and lepton plus missing transverse energy ($\ell+\MET$) final states ($\ell=e,\mu$). The Atlas and CMS collaborations have studied these signatures with LHC data, and have derived quite strong limits on the couplings and masses of the new particles. Resonant dijet and diboson signatures have been analyzed as well, but here we will concentrate on leptonic final states.

Usually, the resonant $\ell^+ \ell^-$ and $\ell+\MET$ signatures are analyzed independently from each other, in terms of the mass, width and couplings of a hypothetical \Zp{} and \Wp{} particle, respectively. On the other hand, each specific theoretical model will predict certain (parameter-dependent) correlations between the properties of \Zp{} and \Wp{} bosons, and thereby between the corresponding neutral and charged signatures. In this paper, we show that some of these correlations are imposed by (spontaneously broken) $\SU{3}_C\times\SU{2}_L\times \U{1}_Y$ gauge invariance, and can be studied from a model-independent point of view. Furthermore, we take advantage of this information to optimize the searches of these extra vectors. Taking into account the $\ell^+ \ell^-$ and $\ell+\MET$ data simultaneously in a joint analysis, we will be able to put stronger limits than the ones obtained from separate analyses.
 
To achieve all this, we work within a general gauge-invariant effective field theory that includes the SM degrees of freedom and the extra vector fields that could give rise to the signatures we are interested in. The principle of gauge invariance strongly restricts the quantum numbers and the interactions of the new vector bosons. In particular, we will argue below that $\ell^+ \ell^-$ resonances at the LHC can only be efficiently produced by the exchange of colorless neutral vector bosons that belong to either singlets or triplets of the isospin $\SU{2}_L$ group, whereas $\ell \nu$ resonances can only be efficiently produced by colorless charged vector bosons in isospin triplets\footnote{Extra scalars can also contribute to both final states, but in this paper we will concentrate on new vector bosons.}. Therefore, our effective theory will simply be an extension of the SM with colorless zero-hypercharge isosinglets ($\calB$) and isotriplet ($\calW$) fields. We will often refer to the fields in these representations simply as singlets and triplets.
	
An important consequence of the allowed vector quantum numbers can be derived straight away: a signature beyond the SM in the $\ell \nu$ final state is always accompanied by a $\ell^+ \ell^-$ signature. Indeed, the pair of charged bosons in a given triplet $\calW$ comes together with a neutral vector boson. Because the charged and neutral extra vectors in the multiplet couple exclusively to left-handed fermions (with couplings of the same size), we shall call them \WpL{} and \ZpL, respectively. The \ZpL{} couples proportionally to the third component of the isospin. On the other hand, the \WpL{} and \ZpL{} bosons are quasi-degenerate. This tight structure implies a complete correlation between the searches of \WpL{} and their neutral partners. Note that, in addition to a \ZpL{} for each pair of \WpL{}, there may exist additional neutral vectors, belonging to the singlets $\calB$, with independent couplings and masses.

This simple picture may get some corrections if the neutral singlets mix with the \ZpL{} bosons, which is possible upon electroweak symmetry breaking. The mixing has two effects: the couplings of the eigenvectors are modified and the mass degeneracy of the triplet is removed. However, the mass of one of the eigenstates always stays close to the \WpL{} mass, as long as the theory is in a perturbative regime and the new particles are heavy. On the other hand, when the mixing is large, the mass of the neutral mass eigenstates must be similar. In this case, the interference effects in the neutral channel must be taken into account. We will show, nevertheless, that in most cases the data can still be interpreted in terms of a single effective resonance.

A mixture of extra triplet and singlet vector fields is interesting for an additional reason: it is the unique way of constructing with extra vectors a sequential \Zp{} model without breaking gauge invariance. Therefore,  {\em any sequential} \Zp{} {\em is always accompanied by another} \Zp{} {\em and by a} \WpL{} \cite{PerezVictoria:2011uj}. The same holds for any \Zp{} boson with significant isospin breaking, i.e.\ with large differences between its couplings to the upper and lower components of the left-handed fermion doublets. The \ZpL{} couplings are recovered in the limit in which the singlet decouples. The sequential \Zp{} (\Wp{}) boson is often employed by the Atlas and CMS collaborations as a benchmark for the interpretation of their dilepton ( $\ell+\MET$) analyses. In order to put these benchmarks on firm theoretical ground, we study here a minimal model that gives rise to sequential \Zp{} and \Wp{} bosons, plus, unavoidably, a sequential extra photon $\gamma^\prime$.  The two heavy neutral states turn out to be nearly degenerate, so their interference cannot be neglected\footnote{This interference is a generic feature of a sequential \Zp{}. It has been studied before in the context of extra-dimensional models~\cite{Rizzo:2009pu,Rizzo:1999en,Bella:2010vi,Bella:2010sc}.}. Therefore, promoting the phenomenological benchmarks into a well defined model has several non-trivial implications: there are correlations between charged and neutral channels and the interference in the neutral channel modifies the cross sections.

Apart from the tight bounds from direct searches, extra vectors with leptonic couplings are also strongly 
constrained by their indirect effects on electroweak precision data (EWPD)\footnote{These limits can be relaxed if a cancellation between the effects of several particles takes place. This, however, requires complex scenarios designed specifically for that purpose \cite{delAguila:2010mx,delAguila:2011yd}.}. 
These limits provide complementary
information to those obtained from direct searches. In particular, given the different dependence of the electroweak constraints on the parameters of the theory, EWPD are still sensitive to regions in the parameter space not yet accessible at the LHC. We will also comment on these bounds and the interplay between them and those from direct searches.

In the next section we briefly review all the possible vector bosons that can couple to leptons. We focus our attention on those cases where such particles can produce sizable leptonic signals at the LHC. As mentioned above, this reduces the spectra to colorless zero-hypercharge isosinglets and isotriplets. These are studied in more detail in Sections \ref{sec:singlet}, \ref{sec:triplet} and \ref{sec:tripletsinglet}. We also present in those sections the electroweak constraints on the different extensions. As an example of a scenario including all these particles, we study in Section \ref{sequential} a simple two-site model generalizing the usual sequential SM. Section \ref{sec:limits} is devoted to the phenomenological analysis of the correlations between
the signals in the $\ell^+ \ell^-$ and $\ell+\MET$ channels in direct searches, for SM extensions including at least one triplet. In particular, we show the improvement on the existing limits once such correlations are taken into account, and compare with the existing electroweak limits. Section \ref{sec:conclusions} presents our conclusions. Finally, we have included one appendix where we develop an effective narrow width approximation to deal with several vector bosons, when these have similar masses and can interfere.

\section{New vector bosons with leptonic couplings}
\label{sec:leptonvectors}

New vector bosons are usually classified according to their quantum numbers under the unbroken gauge symmetries, i.e.\ color and electric charge. Thus, for color singlets we can distinguish between neutral vector bosons Z$^\prime$, charge $\pm 1$ vector bosons W$^\prime$ and vectors with other integer or fractional charges. On the other hand, it is useful to take advantage of the fact that, in the electroweak symmetric phase, the complete theory including the new vectors should respect the full gauge invariance of the SM. Therefore, the extra vectors must furnish complete representations of the group $H=\SU{3}_C \times \SU{2}_L \times \U{1}_Y$, which we will write as $(r_C,r_L)_Y$, with $r_{C,L}$ the dimensionality of the color and isospin representations and $Y$ the hypercharge. Note that in our effective approach we do not require gauge invariance beyond the one of the SM, even if the vector bosons we are considering could be the gauge bosons of an extra gauge group. 
The different irreducible representations of vector fields that can have dimension-four interactions with the SM have been classified and studied in Ref.~\cite{delAguila:2010mx}. Their possible couplings are strongly restricted by gauge invariance. We consider only interactions described by operators with scaling dimension four because they give the leading effects. Finally, the masses of the different components in each multiplet must be equal, up to symmetry breaking effects that come from their coupling to the Higgs doublet and the SM gauge bosons. For large gauge-invariant masses $M$, the splittings in the physical masses squared are always suppressed by $v^2/M^2$, where $v \simeq 246$~GeV is the vacuum expectation value of the Higgs field.

Vector bosons can be produced at the LHC by Drell-Yan or gauge-boson fusion, which require couplings to quarks and gauge bosons, respectively. We are interested in their decay into two leptons: either two charged leptons ($\ell^+ \ell^-$), or a charged lepton and a neutrino ($\ell \nu$). For this, we obviously need vector fields that couple to leptons. We can distinguish two kinds of couplings of the extra vector bosons to the SM fermions. First, there are couplings from interactions that exist already in the gauge-invariant phase. Let us call them direct couplings. 
Second, there are couplings that are not present in the symmetric phase, but arise upon electroweak breaking from the mixing with the SM gauge bosons~\footnote{In this paper we are neglecting the effects of possible extra fermions in the theory, including their mixing with the SM.}. We call them indirect couplings. Assuming perturbative couplings and large masses $M$, the indirect couplings are suppressed by the mixing $s_\alpha$, which is again proportional to $v^2/M^2$. There are also constraints on $s_\alpha$ from electroweak precision tests, which we discuss below. 

It is easy to see that dimension-four direct couplings to leptons can only exist for vector bosons with quantum numbers $(1,1)_0$, $(1,3)_0$ and $(1,2)_{-3/2}$.  On the other hand, mixing, and thus indirect couplings, are only possible for vectors in the representations $(1,1)_0$,  $(1,1)_1$, $(1,3)_0$ and $(1,3)_1$. Therefore, there are only five types of vector fields that can give resonant leptonic signals. Of these, the vectors in the representation $(1,2)_{-3/2}$ can be easily discarded because they do not interact with quarks or gauge bosons, so they cannot be produced at hadron colliders. Let us discuss the other four possibilities in turn.

\begin{itemize}

\item $\calB \in (1,1)_0$. This field is not charged under the SM gauge group $H$. It appears in many extensions of the SM gauge group: if the extended group has the product form $G \times X$, with $H$ a subgroup of $G$, then all the gauge bosons of the factor $X$ belong to this representation of $H$. The only component of this real field is neutral. It can have direct couplings to quarks and leptons, and also indirect couplings via its mixing with the Z boson. Therefore, it can be produced at hadron colliders and it can decay into $\ell^+ \ell^-$. We review it in Section~\ref{sec:singlet}.

\item $\calB^1 \in (1,1)_1$. This complex field contains a pair of \Wp{} bosons of charge $\pm 1$. It is usually called \WpR, since it couples (directly) to right-handed fermions. It appears, for instance, in left-right models (together with a singlet $\calB$). The field $\calB^1$ has direct couplings to quarks, so it can be produced by the Drell-Yan mechanism, but no direct couplings to the SM leptons. On the other hand, it can mix with the SM W bosons and acquire indirect couplings to leptons. This mixing is proportional to the Higgs coupling, $g_{{\cal B}^1}^\phi$, and gives a negative contribution to the $\rho$ parameter. Electroweak precision tests (with the recent value of the Higgs mass) place a strong limit $|g_{{\cal B}^1}^\phi/M_{{\cal B}^1}| < 0.09$ at 95\% C.L. \cite{delAguila:2010mx}. The mixing is also required to be small by data in $K^0-\bar{K^0}$ mixing~\cite{Langacker:1989xa}. The precision-test bounds can be relaxed including extra (leptophobic)  \Zp{} bosons, such that the different contributions to the $\rho$ parameter cancel out. This cancelation could be enforced by custodial symmetry. At any rate, as explained above, the mixing will always be small if the new vectors are heavy and the theory is in the perturbative regime. Moreover, the same mixing allows the \WpR\ to decay into longitudinal gauge bosons and the Higgs, and this channel is kinematically enhanced\footnote{This decay mode, as well as the decay into quarks (which is always present if there is Drell-Yan production), have been studied in detail in~\cite{Grojean:2011vu}.}. From all this, we conclude that the branching ratio into leptons will be small, so this vector boson cannot give an observable contribution to $pp \to \ell \nu$, with $\nu$ a SM neutrino. 

However, this does not preclude the possibility of a $\ell+\MET$ signal : if very light right-handed neutrinos existed, the W$_R^{\prime~\!\pm}$ bosons could decay into a lepton and a right-handed neutrino, which would decay out of the detector\footnote{
On the other hand, having not-so-light right-handed neutrinos results in signals different from those we are interested in \cite{delAguila:2009bb}.}. For the rest of the paper we assume that there are no extra light fermions, and thus we do not consider this possibility further.

\item $\calW \in (1,3)_0$. This multiplet contains a charge $\pm 1$ pair of vector bosons and a neutral one. It appears in extra-dimensional and little Higgs theories. The charged components are often called W$_L^{\prime~\!\pm}$, for they only interact with left-handed fermions. They couple directly to both quarks and leptons. Therefore, the charged components can give observable $\ell+\MET$ signatures at the LHC. The neutral vector boson also couples to left-handed fermions, proportionally to their third component of isospin. We call it Z$_L^\prime$. This neutral vector will contribute to $pp\to \ell^+ \ell^-$ processes. Mixing and indirect couplings are also possible, as we discuss in detail in Section~\ref{sec:triplet}.

\item $\calW^1 \in (1,3)_1$. This less known multiplet contains two neutral vector bosons, a pair with charge $\pm1$ and a pair with charge $\pm 2$. It  has no direct couplings to fermions, but can get indirect couplings from its mixing with the Z and W bosons. It contributes positively to the $\rho$ parameter. Assuming the recently discovered Higgs boson has SM couplings, the electroweak precision data put a limit on the Higgs coupling $|g_{{\cal W}^1}^\phi/M_{{\cal W}^1}| < 0.3$ at the 95\% C.L. ~\cite{delAguila:2010mx}, strongly constraining the mixing with the SM gauge bosons.  Therefore, the production of these vector fields is very small.

\end{itemize}

From this simple analysis we conclude that, in the absence of extra light fermions, sizable $\ell+\MET$ signals at the LHC can only be produced by the charged components of triplet vector bosons $\calW$, whereas sizable $\ell^+ \ell^-$ signals can be produced by either the neutral components in these multiplets or by singlets $\calB$. 
In the following we study in turn SM extensions with extra singlet vector fields, with extra triplet vector fields, and with both singlets and triplets at the same time.

\section{Singlets}
\label{sec:singlet}

The extra singlet vector fields $\calB$ are very well known, especially in the context of abelian extensions of the gauge group. We refer the reader, for instance, to the review in Ref.~\cite{Langacker:2008yv}. Let us stress again that there also exist neutral colorless vector bosons in other multiplets, so it is worthwhile keeping the distinction between the electrically neutral \Zp{} and the field $\calB$ in the $(1,1)_0$ irreducible representation. We will consider a SM extension with only one $\calB$, as the generalization to several singlets is straightforward and would just make the discussion more cumbersome.

The most general gauge-invariant dimension-four Lagrangian containing operators formed  as products of the SM fields and at most two $\calB$ fields reads
\beq
\calL=\calL_\mathrm{SM} + \calL_\calB^0+\calL_\calB^\mathrm{int},
\eeq
where $\calL_\mathrm{SM}$ is the SM Lagrangian, 
\beq
\calL_\calB^0 = -\half \partial_\mu \calB_\nu \partial^\mu \calB^\nu + \half \partial_\mu \calB_\nu \partial^\nu \calB^\mu + \frac{\mu^2_\calB}{2} \calB_\mu \calB^\mu  \label{LagB0} 
\eeq
contains the invariant quadratic terms of the singlet and
\begin{align}
\calL_\calB^\mathrm{int} =& g_{\mathcal{BB}}\calB_{\mu}\calB^{\mu}\phi ^{\dagger} \phi  - \left(i g_{\calB}^\phi \calB^{\mu} \phi ^{\dagger}D_{\mu}\phi  + \mathrm{h.c.} \right)   -  (g^l_\calB)_{ij} \calB^\mu \bar{l_i}\gamma_\mu l_j
- (g^e_\calB)_{ij} \calB^\mu \bar{e_i}\gamma_\mu e_j \nn
& \!-  ( g^q_\calB)_{ij} \calB^\mu \bar{q_i}\gamma_\mu q_j 
- ( g^u_\calB)_{ij} \calB^\mu \bar{u_i}\gamma_\mu u_j - ( g^d_\calB)_{ij} \calB^\mu \bar{d_i}\gamma_\mu d_j
\label{LBint}
\end{align}
describes the interactions of the vector singlet with the SM matter fields. As mentioned above, we have not considered terms with more than two new vectors. Such interactions have a negligible effect for the analysis of leptonic signals or in electroweak precision observables. Thus, they are irrelevant for our analyses.  In writing Eq. (\ref{LBint}) we have also omitted redundant interactions like $\calB^\mu D^\nu B_{\mu\nu}$. These induce a kinetic mixing of the new vectors with the SM gauge boson $B_\mu$ but can be eliminated by a redefinition of fields and couplings. We are using SM-covariant derivatives,
\beq
D_\mu X = (\partial_\mu + ig W_\mu^a T_a + i g^\prime Y B_\mu)\, X \, ,
\eeq
with $T_a$ and $Y$ the $\SU{2}_L$ generators and the hypercharge in the representation of the field $X$, and $g$, $g^\prime$ the SM gauge couplings. The symbols $l_i$ and $q_i$ denote, respectively, the left-handed lepton and quark doublets, while $e_i$, $u_i$ and $d_i$ denote the right-handed singlets. Latin indices $i,j=1,2,3$ label the different fermion generations. In the following we always assume diagonal and universal coupling matrices (in the gauge eingenstate basis). For the first two families, this is actually required to avoid flavor changing neutral currents. The assumption for the third family, made for simplicity, is not essential in our analysis, and it could be easily withdrawn.

\begin{figure}[t!]
\begingroup
  \makeatletter
  \providecommand\color[2][]{%
    \GenericError{(gnuplot) \space\space\space\@spaces}{%
      Package color not loaded in conjunction with
      terminal option `colourtext'%
    }{See the gnuplot documentation for explanation.%
    }{Either use 'blacktext' in gnuplot or load the package
      color.sty in LaTeX.}%
    \renewcommand\color[2][]{}%
  }%
  \providecommand\includegraphics[2][]{%
    \GenericError{(gnuplot) \space\space\space\@spaces}{%
      Package graphicx or graphics not loaded%
    }{See the gnuplot documentation for explanation.%
    }{The gnuplot epslatex terminal needs graphicx.sty or graphics.sty.}%
    \renewcommand\includegraphics[2][]{}%
  }%
  \providecommand\rotatebox[2]{#2}%
  \@ifundefined{ifGPcolor}{%
    \newif\ifGPcolor
    \GPcolortrue
  }{}%
  \@ifundefined{ifGPblacktext}{%
    \newif\ifGPblacktext
    \GPblacktexttrue
  }{}%
  \let\gplgaddtomacro\g@addto@macro
  \gdef\gplbacktexta{}%
  \gdef\gplfronttexta{}%
  \gdef\gplbacktextb{}%
  \gdef\gplfronttextb{}%
  \makeatother
  \ifGPblacktext
    \def\colorrgb#1{}%
    \def\colorgray#1{}%
  \else
    \ifGPcolor
      \def\colorrgb#1{\color[rgb]{#1}}%
      \def\colorgray#1{\color[gray]{#1}}%
      \expandafter\def\csname LTw\endcsname{\color{white}}%
      \expandafter\def\csname LTb\endcsname{\color{black}}%
      \expandafter\def\csname LTa\endcsname{\color{black}}%
      \expandafter\def\csname LT0\endcsname{\color[rgb]{1,0,0}}%
      \expandafter\def\csname LT1\endcsname{\color[rgb]{0,1,0}}%
      \expandafter\def\csname LT2\endcsname{\color[rgb]{0,0,1}}%
      \expandafter\def\csname LT3\endcsname{\color[rgb]{1,0,1}}%
      \expandafter\def\csname LT4\endcsname{\color[rgb]{0,1,1}}%
      \expandafter\def\csname LT5\endcsname{\color[rgb]{1,1,0}}%
      \expandafter\def\csname LT6\endcsname{\color[rgb]{0,0,0}}%
      \expandafter\def\csname LT7\endcsname{\color[rgb]{1,0.3,0}}%
      \expandafter\def\csname LT8\endcsname{\color[rgb]{0.5,0.5,0.5}}%
    \else
      \def\colorrgb#1{\color{black}}%
      \def\colorgray#1{\color[gray]{#1}}%
      \expandafter\def\csname LTw\endcsname{\color{white}}%
      \expandafter\def\csname LTb\endcsname{\color{black}}%
      \expandafter\def\csname LTa\endcsname{\color{black}}%
      \expandafter\def\csname LT0\endcsname{\color{black}}%
      \expandafter\def\csname LT1\endcsname{\color{black}}%
      \expandafter\def\csname LT2\endcsname{\color{black}}%
      \expandafter\def\csname LT3\endcsname{\color{black}}%
      \expandafter\def\csname LT4\endcsname{\color{black}}%
      \expandafter\def\csname LT5\endcsname{\color{black}}%
      \expandafter\def\csname LT6\endcsname{\color{black}}%
      \expandafter\def\csname LT7\endcsname{\color{black}}%
      \expandafter\def\csname LT8\endcsname{\color{black}}%
    \fi
  \fi
  \begin{tabular}{c c}
  \setlength{\unitlength}{0.0350bp}%
  \begin{picture}(7200.00,5040.00)(500,0)%
    \gplgaddtomacro\gplbacktexta{%
      \csname LTb\endcsname%
      \put(1308,1587){\makebox(0,0)[r]{\strut{}-0.4}}%
      \put(1308,2242){\makebox(0,0)[r]{\strut{}-0.2}}%
      \put(1308,2898){\makebox(0,0)[r]{\strut{} 0}}%
      \put(1308,3553){\makebox(0,0)[r]{\strut{} 0.2}}%
      \put(1308,4208){\makebox(0,0)[r]{\strut{} 0.4}}%
      \put(1908,1039){\makebox(0,0){\strut{}-0.4}}%
      \put(2844,1039){\makebox(0,0){\strut{}-0.2}}%
      \put(3779,1039){\makebox(0,0){\strut{} 0}}%
      \put(4715,1039){\makebox(0,0){\strut{} 0.2}}%
      \put(5651,1039){\makebox(0,0){\strut{} 0.4}}%
      \put(538,2897){\rotatebox{-270}{\makebox(0,0){\strut{}$G_{\cal B}^l$[TeV$^{-1}$]}}}%
      \put(3779,709){\makebox(0,0){\strut{}$G_{\cal B}^\phi$[TeV$^{-1}$]}}%
    }%
    \gplgaddtomacro\gplfronttexta{%
    }%
    \gplgaddtomacro\gplbacktexta{%
    }%
    \gplgaddtomacro\gplfronttexta{%
    }%
    \gplgaddtomacro\gplbacktexta{%
    }%
    \gplgaddtomacro\gplfronttexta{%
    }%
    \gplgaddtomacro\gplbacktexta{%
    }%
    \gplgaddtomacro\gplfronttexta{%
    }%
    \gplgaddtomacro\gplbacktexta{%
    }%
    \gplgaddtomacro\gplfronttexta{%
    }%
    \gplbacktexta
    \put(0,0){\includegraphics[scale=0.7]{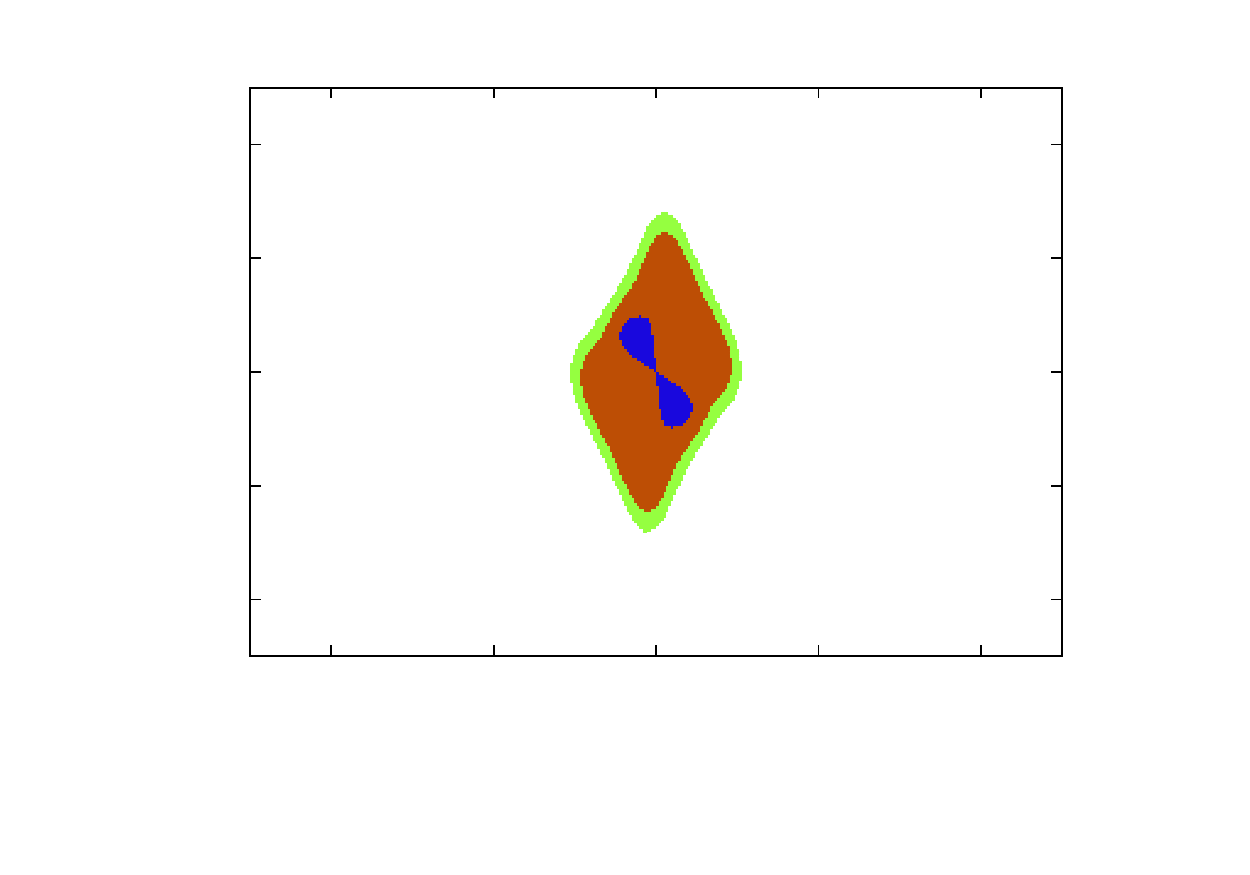}}%
    \gplfronttexta
  \end{picture}%
  &
  \setlength{\unitlength}{0.0350bp}%
  \begin{picture}(7200.00,5040.00)(1400,0)%
    \gplgaddtomacro\gplbacktextb{%
      \csname LTb\endcsname%
      \put(1308,1587){\makebox(0,0)[r]{\strut{}-0.4}}%
      \put(1308,2242){\makebox(0,0)[r]{\strut{}-0.2}}%
      \put(1308,2898){\makebox(0,0)[r]{\strut{} 0}}%
      \put(1308,3553){\makebox(0,0)[r]{\strut{} 0.2}}%
      \put(1308,4208){\makebox(0,0)[r]{\strut{} 0.4}}%
      \put(1908,1039){\makebox(0,0){\strut{}-4}}%
      \put(2844,1039){\makebox(0,0){\strut{}-2}}%
      \put(3780,1039){\makebox(0,0){\strut{} 0}}%
      \put(4715,1039){\makebox(0,0){\strut{} 2}}%
      \put(5651,1039){\makebox(0,0){\strut{} 4}}%
      \put(538,2897){\rotatebox{-270}{\makebox(0,0){\strut{}$G_{\cal B}^l$[TeV$^{-1}$]}}}%
      \put(3779,709){\makebox(0,0){\strut{}$G_{\cal B}^q$[TeV$^{-1}$]}}%
    }%
    \gplgaddtomacro\gplfronttextb{%
    }%
    \gplgaddtomacro\gplbacktextb{%
    }%
    \gplgaddtomacro\gplfronttextb{%
    }%
    \gplgaddtomacro\gplbacktextb{%
    }%
    \gplgaddtomacro\gplfronttextb{%
    }%
    \gplgaddtomacro\gplbacktextb{%
    }%
    \gplgaddtomacro\gplfronttextb{%
    }%
    \gplgaddtomacro\gplbacktextb{%
    }%
    \gplgaddtomacro\gplfronttextb{%
    }%
    \gplbacktextb
    \put(0,0){\includegraphics[scale=0.7]{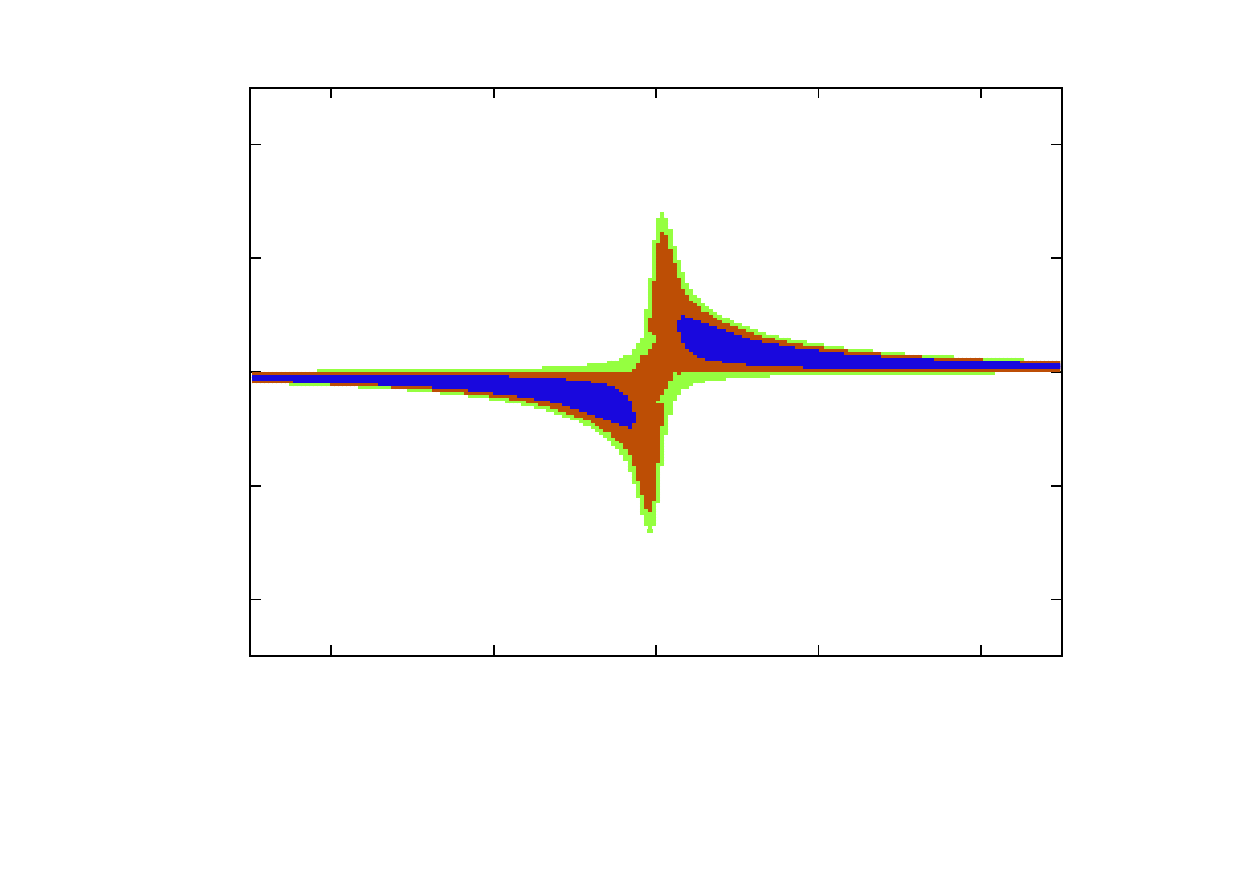}}%
    \gplfronttextb
  \end{picture}%
  \end{tabular}
\endgroup

\caption{Left: From darker to lighter, confidence regions with $\Delta \chi^2\leq2$ (blue), 4 (orange) and 6 ($95\%$ C.L.) (green), respectively, in the $G_{\cal B}^l$-$G_{\cal B}^\phi$ plane for a ${\cal B}$ model with only scalar and left-handed couplings. Right: The same in the $G_{\cal B}^l$-$G_{\cal B}^q$ plane.
}
\label{B0limits}
\end{figure}

In the universal case, the model has six new parameters: the mass $\mu_\calB$, the coupling to the Higgs field $g_\calB^\phi$, and the couplings to the five fermion multiplets, $g_\calB^q$, $g_\calB^u$, $g_\calB^d$, $g_\calB^l$ and $g_\calB^e$. After electroweak breaking, the coupling $g_\calB^\phi$ gives rise to a mixing between the $\calB$ and the Z boson. Electroweak precision data put strong limits on the values of these parameters. We have updated the global fit to electroweak precision observables performed in~\cite{delAguila:2010mx}.\footnote{The fits presented in this paper include the determination of the Higgs mass \cite{:2012gu,:2012gk}, as well as the latest updates in the top \cite{Aaltonen:2012ra,LHCmt} and W masses \cite{Group:2012gb}, and in the weak charge for Cesium \cite{Dzuba:2012kx}. In the SM computations we also consider the two-loop corrections to $R_b$ as described in \cite{Freitas:2012sy}.} We find, at the 95\% C.L. \footnote{Our one-dimensional $95\%$ C.L. EWPD limits are defined by requiring $\Delta \chi^2=3.84$ with respect the minimum value. We assume in the electroweak fits in this paper that all the parameters are real.},
\bea
|G_\calB^\phi|<0.09,&|G_\calB^l|<0.19,&|G_\calB^e|<0.18,
\eea
where $G_{\cal B}^i\equiv g_{\cal B}^i/\mu_{\cal B}$. The quark couplings have very weak bounds, due to the absence of quark to quark processes in precision observables. This can be seen in Figure \ref{B0limits}, where we plot the limits for an SM extension including a ${\cal B}$ with only scalar and left-handed fermionic couplings. On the other hand, the limit on $G_\calB^\phi$ comes partly from the contribution of the singlet to the $\rho$ (or T) parameter.
This limit could be relaxed adding a hypercharged singlet $\calB^1$, which contributes with the opposite sign.

After electroweak symmetry breaking, the singlet vector mixes with the Z gauge boson, with the following mass matrix
\beq
M^2 = \left( \begin{array}{cc}
M_{Z^0}^2 & -\frac{\mathrm{Re}\left[y\right]}{\cos \theta_W} \\
-\frac{\mathrm{Re}\left[y\right]}{\cos \theta_W} & M^2_\calB
\end{array} \right) ,
\eeq
with $M_{Z^0}^2=(g^2+g^{\prime 2}) v^2/4$ the SM Z mass, $\theta_W$ the Weinberg angle, $M^2_\calB = \mu^2_\calB+g_{\calB\calB} v^2$ and $y=g g_\calB^\phi v^2/2$. Up to terms of order $v^4/M_\calB^4$, the eigenvalues of this matrix are
\begin{align}
& M_Z^2 \simeq M_{Z^0}^2 - \frac{y^2}{M_\calB^2 \cos^2 \theta_W}, \nn
& M_{Z^\prime}^2 \simeq M^2_\calB + \frac{y^2}{M_\calB^2 \cos^2 \theta_W} ,
\end{align}
and the matrix is diagonalized by a rotation of angle $\alpha$ with $\sin \alpha \simeq \frac{g \,\mathrm{Re}\left[g_\calB^\phi\right]}{2 \cos \theta_W}\frac{v^2}{M_\calB^2}$.
The negative shift in the mass of the Z boson is responsible for the (positive) contribution to the $\rho$ parameter.

From the electroweak limits above (and, independently, from perturbativity of the couplings) we see that the mixing $\sin \alpha$ is very small. Therefore, we can neglect it for the leptonic signals we are interested in.

In the narrow width approximation, it is possible to study arbitrary \Zp{} bosons (not only the ones from singlets) in terms of only two parameters, called $c_u$ and $c_d$, as proposed in Ref.~\cite{Carena:2004xs} (see also~\cite{Accomando:2010fz} for an application to the LHC). Let us briefly review this formalism.

At the LHC, the cross section for dilepton events mediated by \Zp{} bosons can be written in the narrow width approximation in the following way:
\beq
\sigma(pp\to Z' \to \ell^+\ell^-)=\frac{\pi}{6s}\left[  c_u \omega _u(s,m_{Z'}^2)+ c_d \omega _d (s,m_{Z'}^2) \right]. 
\label{cucdcs}
\eeq
Here, $\omega(s,p^2)$ are model-independent functions that depend on the collision center-of-mass energy $s$ and the dilepton invariant mass (equal to the \Zp{} mass in this approximation), whereas the coefficients $c_u$ and $c_d$ only depend on the couplings of the \Zp:
\beq
c_{q}=\left({g^q_{L}}^2+{g^q_{R}}^2\right) \mathrm{Br} (Z'\to \ell^+\ell^-),
\label{cucd}
\eeq
with $g^{q}_{L,R}$ the left-handed and right-handed couplings of the \Zp{} to the $q$ quark, $q=u,d$. For the case of a \Zp{} from a singlet $\calB$,
\beq
c_{u,d} =\left({g_\calB^{q\phantom{d}}}^2+{g_\calB^{u,d\,}}^2\right)\frac{{g_\calB^l}^2+{g_\calB^{e\phantom{l}}}^2}{3\left(2{g_\calB^{l}}^2+{g_\calB^{e}}^2+6{g_\calB^{q}}^2+3{g_\calB^{u}}^2+3{g_\calB^{d}}^2\right)},
\eeq
where we have assumed only decays into SM fermions, and neglected fermion masses. 
This parametrization is very useful, since the limits in the plane $c_u-c_d$ from the experimental results can be easily compared with the prediction of different models for $c_u$ and $c_d$. This kind of analysis has been done by CMS in their searches of narrow resonances in dileptonic events~\cite{Chatrchyan:2012it}.

\section{Triplets}
\label{sec:triplet}

In this section, we consider an extension of the SM with a vectorial isotriplet $\calW$, in the $(1,3)_0$. This is the simplest scenario with a \WpL{}. The generalization to several extra triplets is straightforward.
The most general dimension-four Lagrangian we can build is given by
\beq
\calL=\calL_\mathrm{SM} + \calL_\calW^0 + \calL_\calW^\mathrm{int},
\eeq
where
\beq
\calL_\calW^0 = -\half [D_\mu \calW_\nu]^a [D^\mu \calW^\nu]_a + \half [D_\mu \calW_\nu]^a [D^\nu \calW^\mu]_a  + \frac{\mu^2_\calW}{2} \calW_\mu^a \calW^\mu_a  \label{LagW0},
\eeq
and
\begin{align}
\calL_\calW^\mathrm{int} =  & g_{\calW\calW} \calW_\mu^a \calW^\mu_a \phi^\dagger \phi - 
(g^l_\calW)_{ij} \calW_a^\mu \bar{l_i}\gamma_\mu \frac{\sigma^a}{2} l_j -( g^q_\calW)_{ij} \calW_a^\mu \bar{q_i}\gamma_\mu \frac{\sigma^a}{2} q_j- \nn
&-\left(i g^\phi_\calW \calW_a^\mu \phi^\dagger \frac{\sigma^a}{2} D_\mu \phi   + \mathrm{h.c.}\right).
\label{LagW1}
\end{align}
We consider again only terms with at most two extra heavy fields, and assume diagonal and universal couplings to the three families of fermions, in the interaction basis. 

\begin{figure}[t!]
\begingroup
  \makeatletter
  \providecommand\color[2][]{%
    \GenericError{(gnuplot) \space\space\space\@spaces}{%
      Package color not loaded in conjunction with
      terminal option `colourtext'%
    }{See the gnuplot documentation for explanation.%
    }{Either use 'blacktext' in gnuplot or load the package
      color.sty in LaTeX.}%
    \renewcommand\color[2][]{}%
  }%
  \providecommand\includegraphics[2][]{%
    \GenericError{(gnuplot) \space\space\space\@spaces}{%
      Package graphicx or graphics not loaded%
    }{See the gnuplot documentation for explanation.%
    }{The gnuplot epslatex terminal needs graphicx.sty or graphics.sty.}%
    \renewcommand\includegraphics[2][]{}%
  }%
  \providecommand\rotatebox[2]{#2}%
  \@ifundefined{ifGPcolor}{%
    \newif\ifGPcolor
    \GPcolortrue
  }{}%
  \@ifundefined{ifGPblacktext}{%
    \newif\ifGPblacktext
    \GPblacktexttrue
  }{}%
  \let\gplgaddtomacro\g@addto@macro
  \gdef\gplbacktexta{}%
  \gdef\gplfronttexta{}%
  \gdef\gplbacktextb{}%
  \gdef\gplfronttextb{}%
  \makeatother
  \ifGPblacktext
    \def\colorrgb#1{}%
    \def\colorgray#1{}%
  \else
    \ifGPcolor
      \def\colorrgb#1{\color[rgb]{#1}}%
      \def\colorgray#1{\color[gray]{#1}}%
      \expandafter\def\csname LTw\endcsname{\color{white}}%
      \expandafter\def\csname LTb\endcsname{\color{black}}%
      \expandafter\def\csname LTa\endcsname{\color{black}}%
      \expandafter\def\csname LT0\endcsname{\color[rgb]{1,0,0}}%
      \expandafter\def\csname LT1\endcsname{\color[rgb]{0,1,0}}%
      \expandafter\def\csname LT2\endcsname{\color[rgb]{0,0,1}}%
      \expandafter\def\csname LT3\endcsname{\color[rgb]{1,0,1}}%
      \expandafter\def\csname LT4\endcsname{\color[rgb]{0,1,1}}%
      \expandafter\def\csname LT5\endcsname{\color[rgb]{1,1,0}}%
      \expandafter\def\csname LT6\endcsname{\color[rgb]{0,0,0}}%
      \expandafter\def\csname LT7\endcsname{\color[rgb]{1,0.3,0}}%
      \expandafter\def\csname LT8\endcsname{\color[rgb]{0.5,0.5,0.5}}%
    \else
      \def\colorrgb#1{\color{black}}%
      \def\colorgray#1{\color[gray]{#1}}%
      \expandafter\def\csname LTw\endcsname{\color{white}}%
      \expandafter\def\csname LTb\endcsname{\color{black}}%
      \expandafter\def\csname LTa\endcsname{\color{black}}%
      \expandafter\def\csname LT0\endcsname{\color{black}}%
      \expandafter\def\csname LT1\endcsname{\color{black}}%
      \expandafter\def\csname LT2\endcsname{\color{black}}%
      \expandafter\def\csname LT3\endcsname{\color{black}}%
      \expandafter\def\csname LT4\endcsname{\color{black}}%
      \expandafter\def\csname LT5\endcsname{\color{black}}%
      \expandafter\def\csname LT6\endcsname{\color{black}}%
      \expandafter\def\csname LT7\endcsname{\color{black}}%
      \expandafter\def\csname LT8\endcsname{\color{black}}%
    \fi
  \fi
  \begin{tabular}{c c}
  \setlength{\unitlength}{0.0350bp}%
  \begin{picture}(7200.00,5040.00)(500,0)%
    \gplgaddtomacro\gplbacktexta{%
      \csname LTb\endcsname%
      \put(1308,1587){\makebox(0,0)[r]{\strut{}-0.4}}%
      \put(1308,2242){\makebox(0,0)[r]{\strut{}-0.2}}%
      \put(1308,2898){\makebox(0,0)[r]{\strut{} 0}}%
      \put(1308,3553){\makebox(0,0)[r]{\strut{} 0.2}}%
      \put(1308,4208){\makebox(0,0)[r]{\strut{} 0.4}}%
      \put(1908,1039){\makebox(0,0){\strut{}-4}}%
      \put(2844,1039){\makebox(0,0){\strut{}-2}}%
      \put(3780,1039){\makebox(0,0){\strut{} 0}}%
      \put(4715,1039){\makebox(0,0){\strut{} 2}}%
      \put(5651,1039){\makebox(0,0){\strut{} 4}}%
      \put(538,2897){\rotatebox{-270}{\makebox(0,0){\strut{}$G_{\cal W}^l$~[TeV$^{-1}$]}}}%
      \put(3779,709){\makebox(0,0){\strut{}$G_{\cal W}^\phi$~[TeV$^{-1}$]}}%
    }%
    \gplgaddtomacro\gplfronttexta{%
    }%
    \gplgaddtomacro\gplbacktexta{%
    }%
    \gplgaddtomacro\gplfronttexta{%
    }%
    \gplgaddtomacro\gplbacktexta{%
    }%
    \gplgaddtomacro\gplfronttexta{%
    }%
    \gplgaddtomacro\gplbacktexta{%
    }%
    \gplgaddtomacro\gplfronttexta{%
    }%
    \gplgaddtomacro\gplbacktexta{%
    }%
    \gplgaddtomacro\gplfronttexta{%
    }%
    \gplbacktexta
    \put(0,0){\includegraphics[scale=0.7]{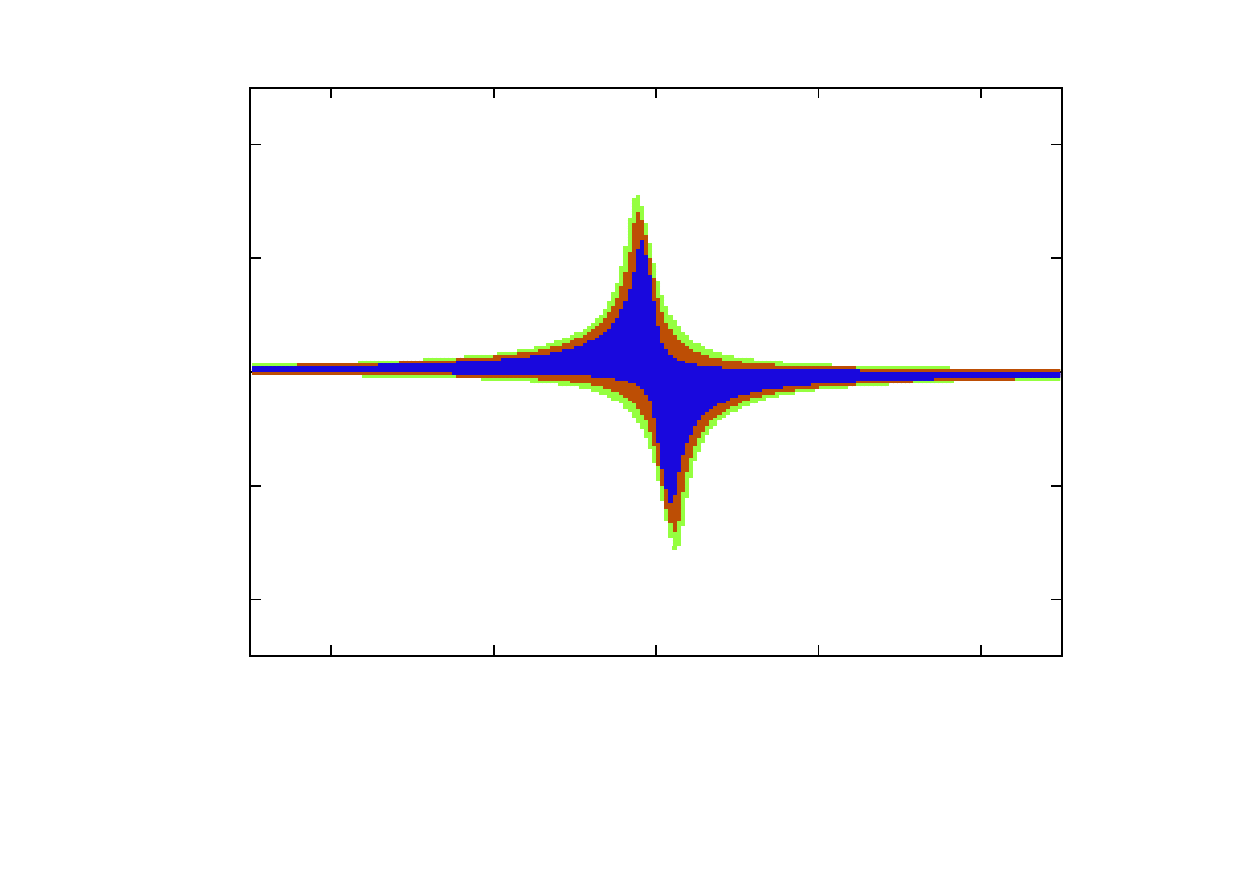}}%
    \gplfronttexta
  \end{picture}%
  &
  \setlength{\unitlength}{0.0350bp}%
  \begin{picture}(7200.00,5040.00)(1400,0)%
    \gplgaddtomacro\gplbacktextb{%
    }%
    \gplgaddtomacro\gplfronttextb{%
    }%
    \gplgaddtomacro\gplbacktextb{%
    }%
    \gplgaddtomacro\gplfronttextb{%
    }%
    \gplgaddtomacro\gplbacktextb{%
    }%
    \gplgaddtomacro\gplfronttextb{%
    }%
    \gplgaddtomacro\gplbacktextb{%
    }%
    \gplgaddtomacro\gplfronttextb{%
    }%
    \gplgaddtomacro\gplbacktextb{%
    }%
    \gplgaddtomacro\gplfronttextb{%
    }%
    \gplgaddtomacro\gplbacktextb{%
      \csname LTb\endcsname%
      \put(1308,1587){\makebox(0,0)[r]{\strut{}-0.4}}%
      \put(1308,2242){\makebox(0,0)[r]{\strut{}-0.2}}%
      \put(1308,2898){\makebox(0,0)[r]{\strut{} 0}}%
      \put(1308,3553){\makebox(0,0)[r]{\strut{} 0.2}}%
      \put(1308,4208){\makebox(0,0)[r]{\strut{} 0.4}}%
      \put(1908,1039){\makebox(0,0){\strut{}-4}}%
      \put(2844,1039){\makebox(0,0){\strut{}-2}}%
      \put(3780,1039){\makebox(0,0){\strut{} 0}}%
      \put(4715,1039){\makebox(0,0){\strut{} 2}}%
      \put(5651,1039){\makebox(0,0){\strut{} 4}}%
      \put(538,2897){\rotatebox{-270}{\makebox(0,0){\strut{}$G_{\cal W}^l$~[TeV$^{-1}$]}}}%
      \put(3779,709){\makebox(0,0){\strut{}$G_{\cal W}^q$~[TeV$^{-1}$]}}%
    }%
    \gplgaddtomacro\gplfronttextb{%
    }%
    \gplbacktextb
    \put(0,0){\includegraphics[scale=0.7]{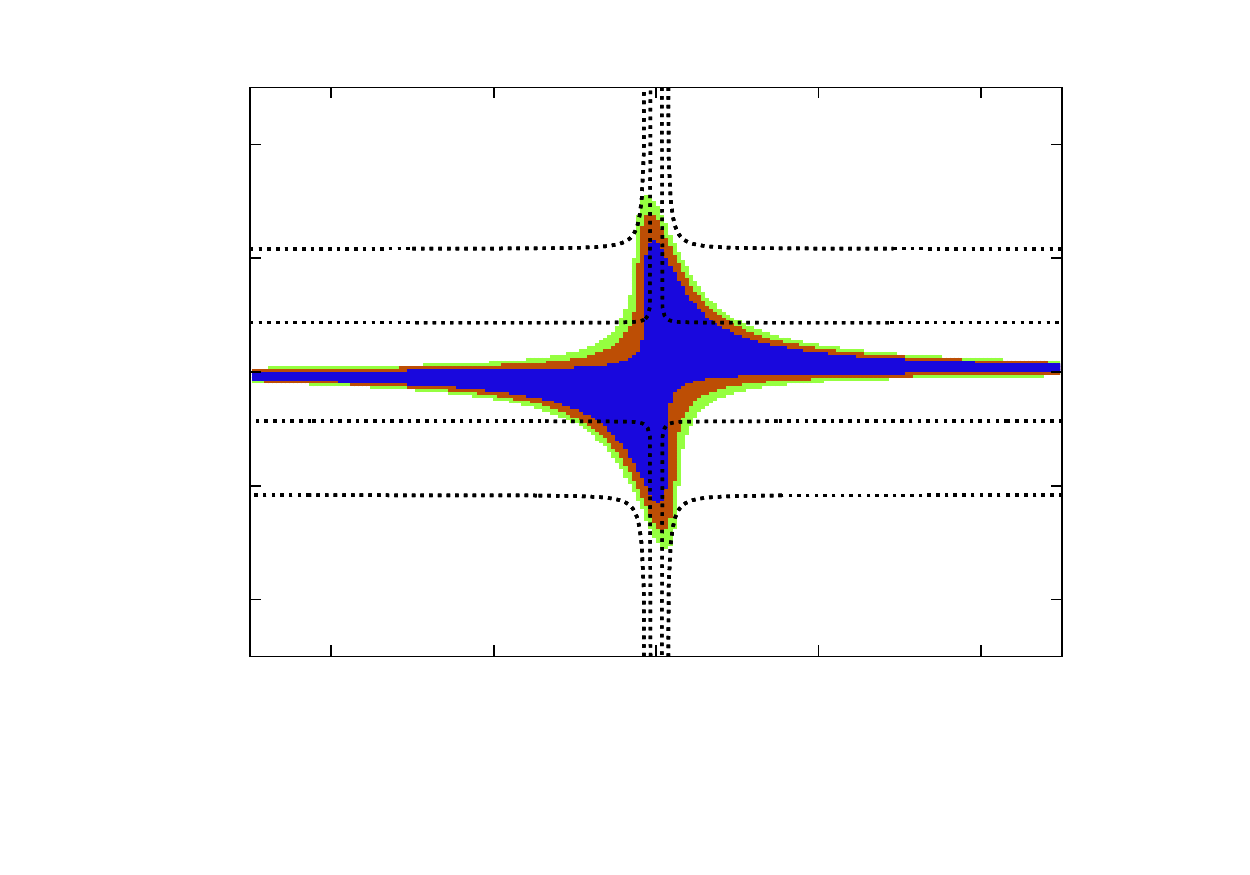}}%
    \gplfronttextb
  \end{picture}%
  \end{tabular}
\endgroup

\caption{Left: From darker to lighter, confidence regions with $\Delta \chi^2\leq2$ (blue), 4 (orange) and 6 ($95\%$ C.L.) (green), respectively, in the $G_{\cal W}^l$-$G_{\cal W}^\phi$ plane. Right: The same in the $G_{\cal W}^l$-$G_{\cal W}^q$ plane. We also plot the curves corresponding to constant values of $|\tilde{g}_{\cal W}/M_{\calW}|=0.1$ (inner curve), 0.25 (outer curve) TeV$^{-1}$. See Eq. (\ref{gW}) and text for details.
}
\label{W0limits}
\end{figure}

Electroweak precision tests put bounds on the coupling to mass ratios $G^{l,q,\phi}_\calW \equiv g^{l,q,\phi}_\calW/M_{\calW}$. These bounds were calculated in \cite{delAguila:2010mx}, and stay mostly unchanged after the updates in the precision data. At $95\%$ C.L. only the leptonic couplings can be bounded, 
\beq
|G_\calW^l|<0.28.
\eeq
The limits are correlated. In particular, for vanishing $g^l_\calW$ there are two directions that are not constrained by the electroweak fit. This is clear in Figure \ref{W0limits}, where we show the limits in two different planes. The flat direction along $g_{\calW}^\phi$ is due to the fact that a vector triplet with real $g_{\calW}^\phi$ preserves custodial symmetry and does not modify the $\rho$ parameter.

After electroweak symmetry breaking, the neutral and charged components of the triplet mix with the Z and the W gauge bosons. The neutral mass matrix is
\beq
M^2_\mathrm{n} = \left( \begin{array}{cc}
M_{Z^0}^2 & \frac{\mathrm{Re}[x]}{\cos \theta_W} \\
\frac{\mathrm{Re}[x]}{\cos \theta_W} & M^2_\calW
\end{array} \right) ,
\eeq
with $M_{Z^0}^2=(g^2+g^{\prime 2}) v^2/4$, $\theta_W$ the Weinberg angle, $M^2_\calW = \mu^2_\calW+g_{\calW\calW} v^2$ and $x=g g_\calW^\phi v^2/4$. Up to terms of order $v^4/M_\calW^4$, the eigenvalues of this matrix are
\begin{align}
& M_Z^2 \simeq M_{Z^0}^2 - \frac{\mathrm{Re}[x]^2}{M_\calW^2 \cos^2 \theta_W}, \nn
& M_{Z^\prime}^2 \simeq M^2_\calW + \frac{\mathrm{Re}[x]^2}{M_\calW^2 \cos^2 \theta_W} ,
\end{align}
and the matrix is diagonalized by a rotation of angle $\alpha_n$ with $\sin \alpha_n \simeq \frac{g  \mathrm{Re}[g^{\phi}_\calW]}{4 \cos \theta_W} \frac{v^2}{M_\calW^2}$.
Similarly, the mass matrix in the charged sector is 
\beq
M^2_\mathrm{c} = \left( \begin{array}{cc}
M_{W^0}^2 & x \\
x^* & M^2_\calW
\end{array} \right) ,
\eeq
with $M_{W^0}^2=g^2 v^2/4$ the SM $W$ mass. Neglecting again terms of order $v^4/M_\calW^4$, the eigenvalues read
\begin{align}
& M_W^2 \simeq M_{W^0}^2 - \frac{|x|^2}{M_\calW^2}, \nn
& M_{W^\prime}^2 \simeq M^2_\calW + \frac{|x|^2}{M_\calW^2}, 
\end{align}
and $\sin \alpha_c \simeq \frac{g |g^{\phi}_\calW|}{4} \frac{v^2}{M_\calW^2}$ is the mixing angle in the $2\times2$ unitary matrix diagonalizing $M_c^2$. It is apparent that, for real $x$, the mixing with the triplet does not spoil the custodial-symmetry relation between the mass of the Z and the W, and at the tree level, $\rho=1$. On the other hand, the neutral and charged heavy vectors are nearly degenerate:
\beq
M_{Z^\prime} \simeq M_{W^\prime}+\frac{v^4}{32M_{W^\prime}^3}\left(g^{\prime 2} \mathrm{Re}[g_\calW^{\phi}]^ 2-g^2 \mathrm{Im}[g_\calW^{\phi}]^2\right) .
\eeq
The splitting of the masses is thus smaller than $0.6\,\mathrm{GeV}$ for $|g_\calW^\phi| < 4\pi$ and \Wp{} masses above 1 TeV. The mixing modifies the interactions of the mass eigenstates with the fermions (including the appearance of a new coupling of the \Zp{} to right-handed singlets) and also induce interactions with the gauge bosons. These effects are suppressed by the mixings $\sin \alpha_{n,c}$, with $\sin \alpha_{n,c}<0.07$ for $|g_\calW^\phi| < 4\pi$ and $M_{W^\prime}>\,1$~TeV.

In order to simplify the LHC analysis we will set $g_\calW^\phi$ to zero, so that the mixing vanishes. This is a good approximation unless the couplings $g^{l,q}_\calW$ are very small (but we need them big to have significant production of the new vectors and significant decay into leptons). 
The parameters of our effective theory are then $M_\calW$, $g^l_\calW$ and $g^q_\calW$. The masses of the new vector bosons are degenerate, $M_{Z^\prime}=M_{W^\prime} =M_\calW$, while their couplings to fermions in the mass eigenstate basis are given by
\begin{align}
& \calL^\mathrm{CC}_\calW =  -\frac{1}{\sqrt{2}} \left(g^q_\calW \bar{u}_{Li}\gamma_\mu V_{ij} d_{Lj} + g^l_\calW  \bar{e}_{Li}\gamma_\mu  \nu_{Li} \right)  W^{\prime +}_\mu + 
\mathrm{h.c.}  ,  \label{chargedcurrent}  \\
& \calL^\mathrm{NC}_\calW =  -\frac{1}{2} \left[ g^q_\calW \left( \bar{u}_{Li}\gamma_\mu u_{Li} - \bar{d}_{Li} \gamma_\mu  d_{Li}  \right)
+ g^l_\calW \left(\bar{\nu}_{Li} \gamma_\mu \nu_{Li} - \bar{e}_{Li} \gamma_\mu e_{Li} \right) \right]  Z^{\prime}_\mu ,\label{tripletZcouplings}
\end{align}
with $V$ the CKM matrix.

When $g^l_\calW=g^q_\calW=g$, in the charged sector we have exactly the sequential \WpL{} model commonly used as a benchmark in the Atlas and CMS analyses of charged vector bosons. On the other hand, the \Zp{} is not sequential. It couples to the third component of isospin. In particular, it only interacts with left-handed fermions.  The isospin dependent couplings of this neutral vector \ZpL{} reveal that it belongs to a multiplet of dimension higher than one.

The neutral vector boson Z$^\prime_L$ can be easily analyzed, in the narrow width approximation, in terms of the $c_u$ and $c_d$ parameters:
\beq
c_u=c_d=\frac{\tilde{g}^2_{\mathcal{W}}}{96},
\eeq
where we define the effective coupling
\beq
\tilde{g}_\calW=\frac{2 g^q_\calW g^l_\calW}{\sqrt{3 {g^q_\calW}^2 + {g^l_\calW}^2}}.
\label{gW}
\eeq
Furthermore, the cross section of the $pp\to W_L^\prime \to \ell\nu$ process also depends, in the narrow width approximation, on the same combination of couplings, $\tilde{g}_\calW$. Therefore, in this approximation, both the $\ell^+ \ell^-$ and $\ell+\MET$ signatures of the triplet depend only on two parameters: the common mass $M_\calW$ and the effective coupling $\tilde{g}_\calW$.
In Figure \ref{W0limits}, where we show the regions allowed by precision tests, we have also plotted contour curves for several values of $|{\tilde g}_{\cal W}/M_{\calW}|$. From this we can read $|{\tilde g}_{\cal W}/M_{\calW}|\lesssim 0.3$ TeV$^{-1}$.

\begin{figure}[t!]
\begin{center}
\input{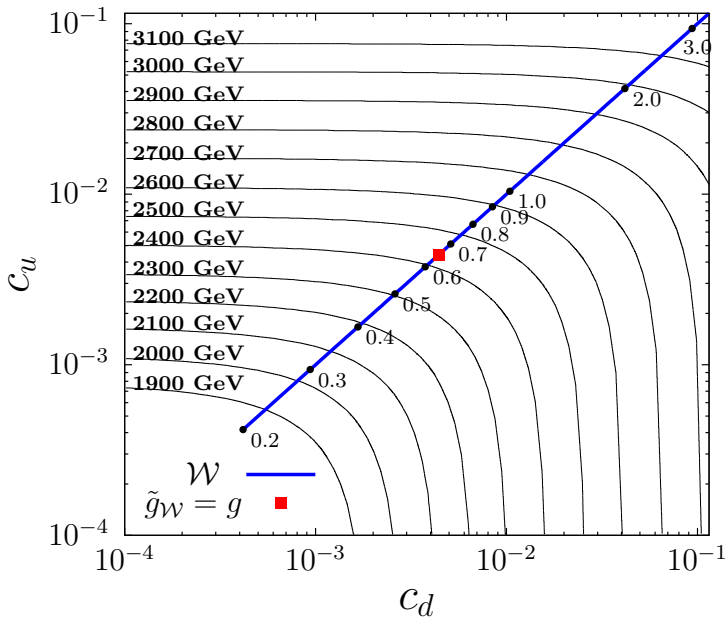}
\caption{$95\%$ C.L. exclusion limits for the cross section of the \Zp{} plotted over the $c_u-c_d$ plane.  From Ref.~\!\cite{Chatrchyan:2012it}. The $c_u$ and $c_d$ values of the triplet as a function of $\tilde{g}_{\mathcal{W}}$ are represented by the blue line.}
\label{CuCdTriplet}
\end{center}
\end{figure}

\section{Triplets and Singlets}
\label{sec:tripletsinglet}

Let us now study the scenario in which we have simultaneously extra isotriplets  $\calW$ in the $(1,3)_0$ representation and extra singlets $\calB$, in the $(1,1)_0$. For definiteness, we consider just one extra vector boson of each kind. The main interest of this extension of the minimal case is the possibility of having large mixing of the \Zp{} inside the triplet with the singlet. The most general gauge-invariant dimension-four Lagrangian we can write is
\beq
\calL=\calL_\mathrm{SM} + \calL_\calW^0 + \calL_\calW^\mathrm{int} + \calL_\calB^0+\calL_\calB^\mathrm{int} + \calL_{\calW\calB} .
\eeq
The new term reads
\beq
\calL_{\calW\calB}= g_{\mathcal{WB}}\calW^{a}_{\mu}\calB^{\mu}\phi ^{\dag} \frac{\sigma_{a}}{2} \phi,
\eeq
while the other ones have already been written in the previous sections.

We have performed a global fit to electroweak precision data of extensions of the SM with a singlet and a triplet. These can be compared to the ones above for each single extension.  Electroweak precision tests require at the $95\%$ C.L. 
\bea
|G_\calB^\phi|<0.11,&|G_\calB^l|<0.15,&|G_\calB^e|<0.11,\label{limite_singlettriplet}\\
&0.08<|G_\calW^l|<0.34, & \nonumber
\eea
where $G_{\cal V}^i\equiv g_{\cal V}^i/\mu_{\cal V}$. As in the case of each individual extension, neither the quark couplings nor the Higgs coupling to $\calW$ can be constrained at this confidence level. The combined fit
is characterized by a significantly deeper minimum than any of the separate extensions (we find $\chi^2-\chi^2_{\mathrm{SM}}=-16.4$, which is 4.1 [15.1] units lower than the fit to the $\calB$ [$\calW$] extension, for 3 [6] extra degrees of freedom), and a preference for slightly smaller (larger) leptonic couplings to the $\calB$ ($\calW$).

Let us study the mass terms. After electroweak breaking the charged vector mass matrix is the same as in the previous section, whereas the one for the neutral vectors $($Z$_\mu,\calW^3_\mu,\calB_\mu)$ is
\beq
M_n^2=
\left(
\begin{array}{c c c}
M_{Z^0}^2 &  \frac{\mathrm{Re}[x]}{\cos \theta_W} & - \frac{\mathrm{Re}[y]}{\cos \theta_W}  \\
\frac{\mathrm{Re}[x]}{\cos \theta_W} & M_{\calW}^2 & -\frac{k}{2} \\
- \frac{\mathrm{Re}[y]}{\cos \theta_W} & -\frac{k}{2}  & M_\calB^2
       \end{array}
\right),
\eeq
where $k=g_{\calW\calB}v^2/2$, $M_\calB^2=\mu_\calB^2+g_{\calB\calB}v^2$ and the other symbols were defined in the previous sections. We also introduce $\Delta_{\calB\calW}=M^2_\calB-M^2_\calW$. We assume that the vector and triplet fields are heavy, $M_\calB, M_\calW \geq 1~\mathrm{TeV}$ and that the couplings are not too large. Then the mixings of the Z boson with the extra bosons are small: $\sin \alpha_{1,2}\sim v^2/M_{\calB,\calW}^2$. Moreover, the coupling $g_\calB^\phi$ is constrained by EWPD, according to \refeq{limite_singlettriplet}.

With these constraints, the impact of the mixing of SM gauge bosons and heavy vectors on the processes we are studying is small. Therefore, we can neglect this mixing and simply set $g_\calB^\phi=g_\calW^\phi=0$. The neutral mass eigenstates are then Z$_\mu$, the physical Z boson, and
\begin{align}
& \mbox{Z}^\prime_{1,\mu}= \cos \theta \, \calW^3_\mu - \sin \theta \, \calB_\mu , \\
& \mbox{Z}^\prime_{2,\mu}=\sin \theta \, \calW^3_\mu + \cos \theta \, \calB_\mu ,
\end{align}
where $\tan 2\theta = \frac{k}{\Delta_{\calB\calW}}$ and $\theta \in (-\frac{\pi}{4},\frac{\pi}{4}]$. Note that, with our definitions, Z$^\prime_{1\mu}$ and Z$^\prime_{2\mu}$ are mostly $\calW_\mu^3$ and $\calB_\mu$ like, respectively. The masses are given by $M_{W}=M_{W^0}$, $M_{Z}=M_{Z^0}$, $M_{W^\prime} = M_\calW$  and
\begin{align}
& M_{Z_1^\prime}^2 = M_\calW^2 - \mathrm{sign}(\Delta_{\calB\calW})  \frac{\sqrt{\Delta_{\calB\calW}^2+k^2}-|\Delta_{\calB\calW}|}{2} \nn
& \phantom{M_{Z_1^\prime}^2} = M_\calW^2 - \Delta_{21} \sin^2\theta  , \\
& M_{Z_2^\prime}^2= M_\calW^2 + \mathrm{sign}(\Delta_{\calB\calW}) \frac{\sqrt{\Delta_{\calB\calW}^2+k^2}+|\Delta_{\calB\calW}|}{2} \nn 
& \phantom{M_{Z_2^\prime}^2}= M_\calW^2+\Delta_{21} \cos^2\theta  ,
\end{align} 
where
\bea
\Delta_{21}&=&  \mathrm{sign}(\Delta_{\calB\calW}) \sqrt{\Delta_{\calB\calW}^2+k^2}  \nn
&=& M_{Z_2^\prime}^2 -M_{Z_1^\prime}^2 .
\eea
In all cases, one \Zp{} is heavier than the \Wp{} and the other \Zp{} is lighter. 
Another useful relation is
\beq
\sin 2\theta = \frac{g_{\calW\calB} v^2}{2\Delta_{21}} . \label{mixingangle}
\eeq

\begin{figure}[t!]
\begingroup
  \makeatletter
  \providecommand\color[2][]{%
    \GenericError{(gnuplot) \space\space\space\@spaces}{%
      Package color not loaded in conjunction with
      terminal option `colourtext'%
    }{See the gnuplot documentation for explanation.%
    }{Either use 'blacktext' in gnuplot or load the package
      color.sty in LaTeX.}%
    \renewcommand\color[2][]{}%
  }%
  \providecommand\includegraphics[2][]{%
    \GenericError{(gnuplot) \space\space\space\@spaces}{%
      Package graphicx or graphics not loaded%
    }{See the gnuplot documentation for explanation.%
    }{The gnuplot epslatex terminal needs graphicx.sty or graphics.sty.}%
    \renewcommand\includegraphics[2][]{}%
  }%
  \providecommand\rotatebox[2]{#2}%
  \@ifundefined{ifGPcolor}{%
    \newif\ifGPcolor
    \GPcolortrue
  }{}%
  \@ifundefined{ifGPblacktext}{%
    \newif\ifGPblacktext
    \GPblacktexttrue
  }{}%
  \let\gplgaddtomacro\g@addto@macro
  \gdef\gplbacktexta{}%
  \gdef\gplfronttexta{}%
  \gdef\gplbacktextb{}%
  \gdef\gplfronttextb{}%
  \makeatother
  \ifGPblacktext
    \def\colorrgb#1{}%
    \def\colorgray#1{}%
  \else
    \ifGPcolor
      \def\colorrgb#1{\color[rgb]{#1}}%
      \def\colorgray#1{\color[gray]{#1}}%
      \expandafter\def\csname LTw\endcsname{\color{white}}%
      \expandafter\def\csname LTb\endcsname{\color{black}}%
      \expandafter\def\csname LTa\endcsname{\color{black}}%
      \expandafter\def\csname LT0\endcsname{\color[rgb]{1,0,0}}%
      \expandafter\def\csname LT1\endcsname{\color[rgb]{0,1,0}}%
      \expandafter\def\csname LT2\endcsname{\color[rgb]{0,0,1}}%
      \expandafter\def\csname LT3\endcsname{\color[rgb]{1,0,1}}%
      \expandafter\def\csname LT4\endcsname{\color[rgb]{0,1,1}}%
      \expandafter\def\csname LT5\endcsname{\color[rgb]{1,1,0}}%
      \expandafter\def\csname LT6\endcsname{\color[rgb]{0,0,0}}%
      \expandafter\def\csname LT7\endcsname{\color[rgb]{1,0.3,0}}%
      \expandafter\def\csname LT8\endcsname{\color[rgb]{0.5,0.5,0.5}}%
    \else
      \def\colorrgb#1{\color{black}}%
      \def\colorgray#1{\color[gray]{#1}}%
      \expandafter\def\csname LTw\endcsname{\color{white}}%
      \expandafter\def\csname LTb\endcsname{\color{black}}%
      \expandafter\def\csname LTa\endcsname{\color{black}}%
      \expandafter\def\csname LT0\endcsname{\color{black}}%
      \expandafter\def\csname LT1\endcsname{\color{black}}%
      \expandafter\def\csname LT2\endcsname{\color{black}}%
      \expandafter\def\csname LT3\endcsname{\color{black}}%
      \expandafter\def\csname LT4\endcsname{\color{black}}%
      \expandafter\def\csname LT5\endcsname{\color{black}}%
      \expandafter\def\csname LT6\endcsname{\color{black}}%
      \expandafter\def\csname LT7\endcsname{\color{black}}%
      \expandafter\def\csname LT8\endcsname{\color{black}}%
    \fi
  \fi
  \begin{tabular}{c c}
  \setlength{\unitlength}{0.0400bp}%
  \begin{picture}(7200.00,5040.00)(1250,0)%
    \gplgaddtomacro\gplbacktextb{%
    }%
    \gplgaddtomacro\gplfronttextb{%
      \csname LTb\endcsname%
      \put(3600,4600){\makebox(0,0){\strut{}Scenario 1: $\left|M_\calB^2-M_\calW^2\right|\gg v^2$}}%
      \put(5000,850){\makebox(0,0){\strut{}W$^\prime$}}%
      \put(5400,525){\makebox(0,0){\strut{}Z$^\prime_1\approx~\!$Z$^\prime_L$}}%
      \put(5000,3880){\makebox(0,0){\strut{}Z$^\prime_2$}}%
    }%
    \gplbacktextb
    \put(0,0){\includegraphics[scale=0.8]{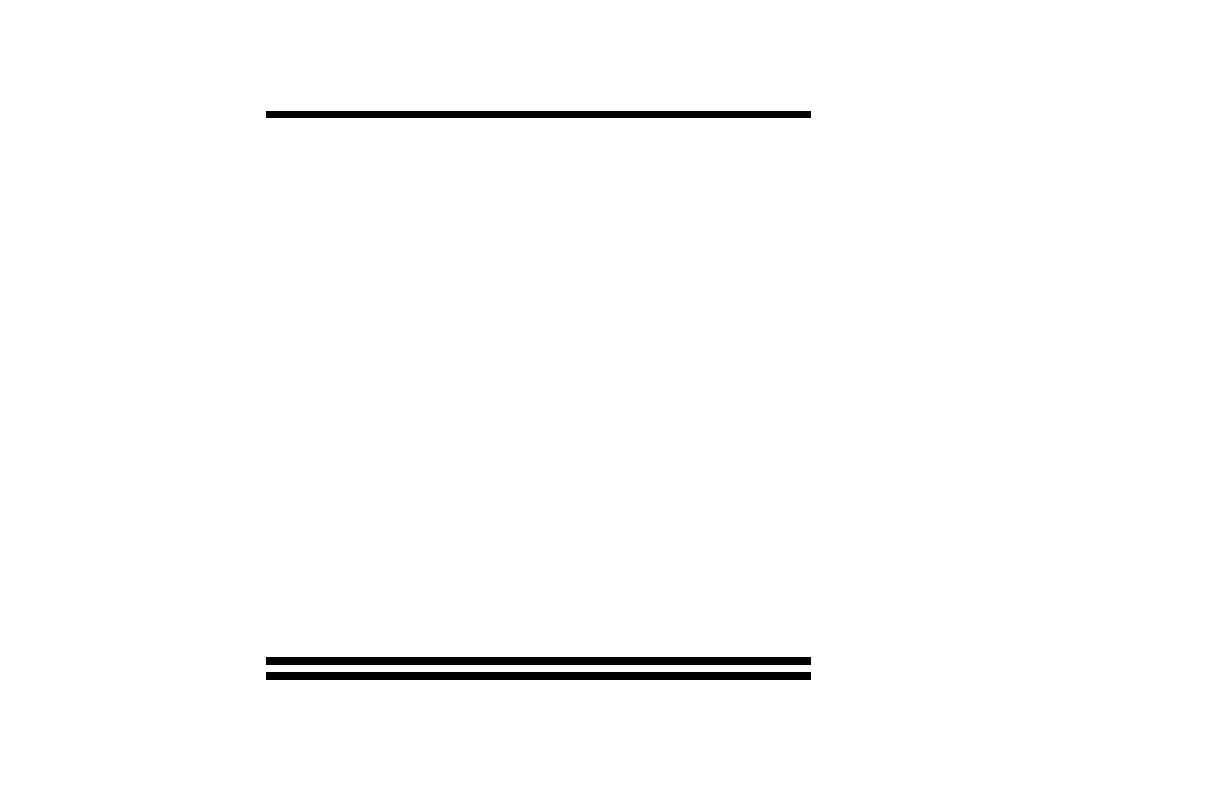}}%
    \gplfronttextb
  \end{picture}%
  &
  \setlength{\unitlength}{0.0400bp}%
  \begin{picture}(7200.00,5040.00)(2400,0)%
    \gplgaddtomacro\gplbacktexta{%
    }%
    \gplgaddtomacro\gplfronttexta{%
      \csname LTb\endcsname%
      \put(3600,4600){\makebox(0,0){\strut{}Scenario 2: $\left|M_\calB^2-M_\calW^2\right|\lesssim v^2$}}%
      \put(5000,2150){\makebox(0,0){\strut{}W$^\prime$}}%
      \put(5000,1700){\makebox(0,0){\strut{}Z$^\prime_1$}}%
      \put(5000,2600){\makebox(0,0){\strut{}Z$^\prime_2$}}%
    }%
    \gplbacktexta
    \put(0,0){\includegraphics[scale=0.8]{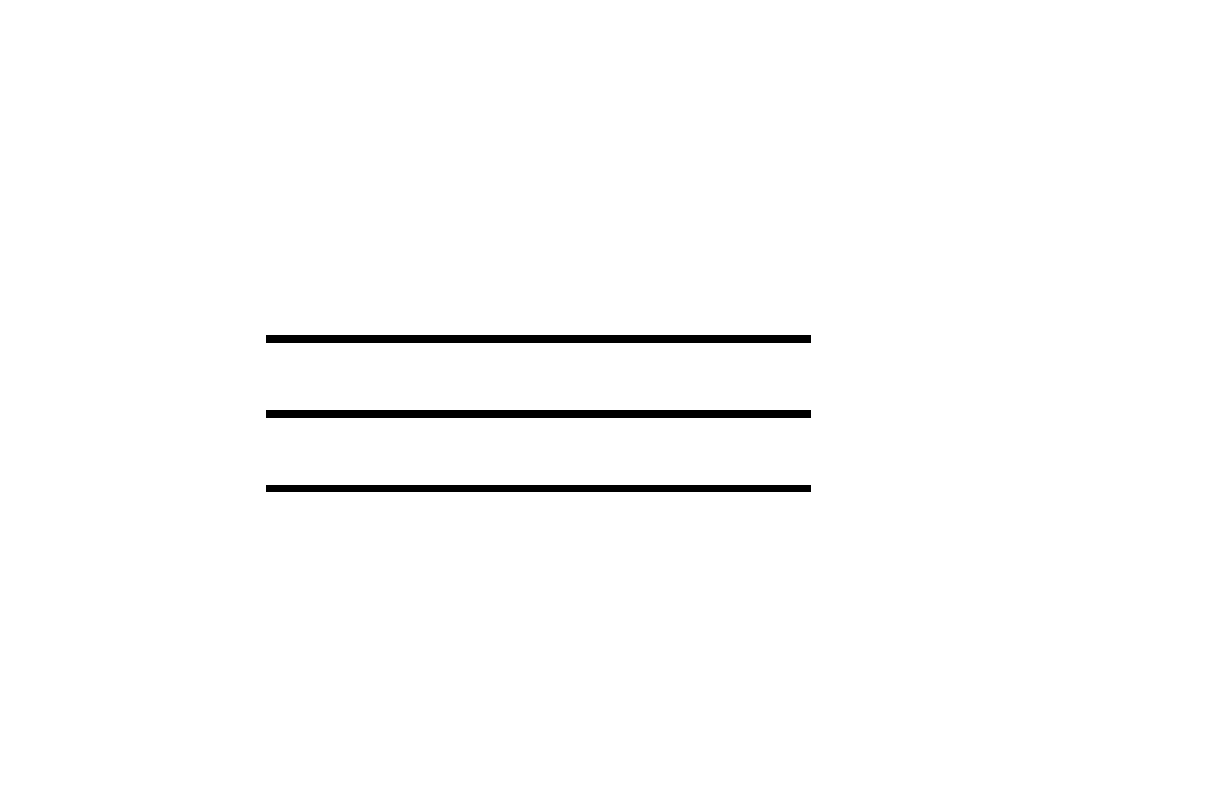}}%
    \gplfronttexta
  \end{picture}%
  \end{tabular}
\endgroup

\caption{Pattern of the Z$^\prime_{1,2}$ and W$^\prime$ masses in the two regimes described in the text.
\label{fig:regimes}}
\label{lim:GenSeq}
\end{figure}

It is important to note that, whereas $|\Delta_{\calB\calW}|$ (and $|\Delta_{21}|$) can in principle be arbitrarily large, $k$ is proportional to $v^2$, and it is thus bounded in a perturbative theory. A direct consequence of this is that the mass of the $\calW^3$-like neutral boson, Z$^\prime_1$ must always be similar to the one of the charged bosons \Wp. Indeed, $|M_{W^\prime}^2-M_{Z_1^\prime}^2|$ is a monotonically decreasing function of $|\Delta_{\calB\calW}|$ and it is bounded from above by $k^2/4$. For $M_{W^\prime} \geq 1\, \mathrm{TeV}$, we find $|M_{W^\prime}-M_{Z_1^\prime}| \leq 94\, \mathrm{GeV}$ if $g_{\calW\calB}\leq 4\pi$ and $|M_{W^\prime}-M_{Z_1^\prime}| \leq 7.8\, \mathrm{GeV}$ if $g_{\calW\calB}\leq 1$. We can distinguish two regimes:
\begin{enumerate}
\item $|\Delta_{21}| \gg v^2$. In this case the mixing angle $\theta$ is small and $M_{Z^\prime_1} \simeq M_{W^\prime}$. The couplings of Z$^\prime_1$ are approximately equal to the ones of a \ZpL. The two neutral bosons can be produced independently from each other. Therefore, the triplet and the singlet can be treated separately in direct searches.
\item $|\Delta_{21}| \lesssim v^2$. In this case $\theta$ is not constrained by perturbativity. The masses of all the heavy vector bosons are similar, with differences of 30~GeV at most for extra bosons above 1~TeV. Therefore, in this regime the interference in the exchange of the two neutral bosons cannot be neglected.
\end{enumerate}
The pattern of masses in the two regimes is illustrated in Figure \ref{fig:regimes}. The second regime requires some fine tuning in the masses, which might be enforced by a symmetry in the fundamental theory.

Let us now consider the interactions with fermions. We assume again that all the couplings to fermions are flavor diagonal and universal, in the interaction basis. Then, the interactions of the charged vector bosons are as for the triplet, \refeq{chargedcurrent}, whereas the interactions of the two heavy neutral bosons read
\beq
\mathcal{L}^\mathrm{NC}_{\calW\calB}  = -\sum_\psi \left\{Z^\prime_{1 \mu} \left( T_3^{\psi} g_{\mathcal{W}}^{\psi}\cos \theta - g_{\mathcal{B}}^{\psi} \sin \theta \right) + Z^\prime_{2 \mu} \left( T_3^{\psi} g_{\mathcal{W}}^{\psi}\sin \theta + g_{\mathcal{B}}^{\psi} \cos \theta \right) \right\} \bar{\psi} \gamma ^{\mu}\psi, 
\label{neutralcurrent}
\eeq
where the sum in $\psi$ is over all chiral fermions, $g_{\calW,\calB}^{\psi}$ are the couplings to $\psi$, and $T_3^{\psi}$ is the third component of isospin of $\psi$. 

Our general effective theory contains ten parameters: $M_\calW$, $M_\calB$, $g_{\calW\calB}$, $g_\calW^{q,l}$, $g_\calB^{q,l,e,u,d}$. The first three can be traded by the mass $M_{W^\prime}$ of the charged heavy vector, the splitting in the mass squared $\Delta_{21}$ of the two heavy neutral bosons, and the mixing angle $\theta$. As stressed above, there is a $\Delta_{21}$-dependent perturbativity limit on the mixing angle, given by \refeq{mixingangle}. When $|\Delta_{21}|\to \infty$, the singlet and the triplet decouple.

As we have already pointed out, in Regime 2 the two neutral states always have similar masses. Thus, the interference of their contributions  to the cross section of the $pp\to \ell^+ \ell^-$ process must be taken into account. Furthermore, the propagation of the two bosons becomes interconnected at the quantum level. Already at one loop, the self-energies mix their masses and widths, and the propagator is a $2\times 2$ complex matrix in the space of the two neutral bosons \cite{Cacciapaglia:2009ic}. When the mass and width matrices do not commute, they cannot be simultaneously diagonalized and the processes with exchange of these particles in the $s$ channel are not well described by a Breit-Wigner propagator for each separate particle. The resummed matrix propagator is the inverse of the renormalized 1-particle-irreducible two-point function. It has the form
\beq
P_{\mu\nu}=\frac{-i}{p^2-\mathcal{M}^2(p^2)}\left[ g_{\mu\nu} -p_{\mu}p_{\nu}\frac{1+\Pi_L(p^2)}{\mathcal{M}^2(p^2)+p^2\Pi_L(p^2)}\right],
\label{Pmunu}
\eeq
where $\frac{A}{B}\equiv A B^{-1}$ and $\mathcal{M}^2(p^2)=M^2+\Pi_T(p^2)$. The matrices $\Pi_{T,L}$ are the transverse and longitudinal parts of the self-energy matrix of the vector bosons, and $M^2$ is the renormalized mass-squared matrix in an on-shell scheme: $\frac{d}{d p^2} \mathcal{M}^2 (\bar{m}^2)=0$ and $M^2=\mathrm{Re}~\!\mathcal{M}^2(\bar{m}^2)$, with the condition $\bar{m}^2=\frac{1}{2} \mathrm{Tr}~\!M^2$. In the processes we are interested in, this propagator is always contracted with currents of light fermions. Therefore, we can safely neglect the longitudinal part. Finally, neglecting also non-local parts in $\mathrm{Re}~\!\mathcal{M}^2(p^2)$, we obtain a generalization of the Breit-Wigner approximation
\beq
P_{\mu\nu}=-i \frac{g_{\mu\nu}}{p^2-M^2-i~\!\mathrm{Im}~\!\Pi_T(p^2)}.
\label{propagator}
\eeq
where it should be remembered that $\Pi_T$ and $M^2$ are matrices. This is the propagator we have used in our exact calculations.

It is quite remarkable that a modified narrow width approximation exists also in this more intricate case.  We prove in the appendix that, when the individual widths are larger than the mass interspace but much smaller than the masses, the cross section can be approximated by \refeq{cucdcs} with
\beq
c_q=  \frac{1}{6\pi\Tr\Sigma} \left[ \mathrm{Tr}\left(G^lG^q\right)+\frac{\mathrm{Tr}\left(G^l \widetilde{\Sigma} G^q \widetilde{\Sigma}\right)}{\det \Sigma} \right],~ q=u,d,
\label{cqint}
\eeq
where we have defined the $2\times2$ matrices
\beq
G^{f}_{ij}=\frac{1}{2} \left[(g_L^{f})_i (g_L^{f})_j+ (g_R^{f})_i (g_R^{f})_j \right] ,
\label{def}
\eeq
with $(g_{L,R}^f)_i$ the chiral couplings of the Z$^\prime_i$ to each SM fermion $f$ (with $g_R^\nu=0$), and
\beq
\Sigma_{ij} = \frac{1}{12\pi} \, \sum_{f=u,d,e,\nu} G^{f}_{ij} .
\eeq
The matrix $\widetilde{\Sigma}$ is the adjugate
\beq
\widetilde{\Sigma}_{ij} = \epsilon_j^m \epsilon_i^n \Sigma_{mn}.
\eeq
In some special models it may happen that $\mathrm{det}\,\Sigma=0$. This means that one linear combination of the two \Zp{} decouples from the SM fermions (or both, but this case is trivial), and the usual expressions for $c_{u,q}$ are recovered. 

\section{The generalized sequential model}
\label{sequential}

In any weakly-coupled model containing a \Zp{} with the same couplings as the SM Z boson, there are triplet and singlet vector fields with sizable mixing. From \refeq{neutralcurrent}, we find $\sin \theta \gtrsim 0.17$ for fermionic couplings smaller than one (of both heavy neutral bosons). Using \refeq{mixingangle} and $g_{\calW\calB}\leq 1$, this requires $\Delta_{21}\lesssim 0.09 \,\mathrm{TeV}^2$. Therefore, along with the sequential \Zp{} there is another \Zp{} and a \Wp{}, with masses differing by less than 50 GeV if they are heavier than 1~TeV. This implies that the sequential \Zp{} cannot be studied independently from, at least, the additional \Zp. If the \Wp{} is also required to have the same couplings as the SM W boson, it is easy to see that the second neutral vector will couple to the electric charge, exactly as the photon. In the following we propose a minimal model that implements a generalization of such a sequential SM with an arbitrary mixing angle. 

We will concentrate on the electroweak sector. The model is a two-site moose with the SM electroweak group on each site\footnote{This model is an extreme deconstruction of a theory with the electroweak gauge bosons in an extra dimension. For certain values of the parameters, it can also be viewed as a simplified representation of an elementary sector coupled to a strongly-coupled one, as proposed in~\cite{Contino:2006nn}. Moreover, the same gauge group is used in the Littlest Higgs model~\cite{ArkaniHamed:2002qy}.}. The gauge group is thus $\SU{2}_1\times \U{1}_1 \times \SU{2}_2 \times \U{1}_2$. Let $g_i$ ($g^\prime_i$) and $W_\mu^i$ ($B_\mu^i$) be, respectively, the gauge coupling and gauge boson of the factor $\SU{2}_i$ ($\U{1}_i$). A scalar field $\chi$ in the representation $(2_Y,2_Y)$ of this group takes a vacuum expectation value 
\beq
\langle \chi \rangle = \frac{\Lambda}{\sqrt{2}}  \left(\begin{array}{cc}1 & 0 \\0 & 1\end{array}\right),
\eeq
which breaks the gauge symmetry to the diagonal subgroup $\SU{2}_L\times \U{1}_Y$, with gauge couplings $g$, $g^\prime$ satisfying $g^{-2}=g_1^{-2}+g_2^{-2}$, $g^{\prime -2}=g_1^{\prime -2}+g_2^{\prime -2}$. We assume that the excitations of $\chi$ around its vacuum are heavy enough to be ignored in our analyses. In addition, we introduce the SM matter content in SM representations, but with the fermions and the Higgs field transforming only under the gauge group in one site. There are two possibilities: 
\begin{itemize}
\item[A.] The fermions and the Higgs field "live'' on the same site (which we call site~1).
\item[B.] They "live" on opposite sites (fermions on site~1 and the Higgs  doublet on site~2).
\end{itemize}
In the electroweak symmetric phase, the mass eigenstate vector fields are
\begin{align}
& W_\mu=\cos \varphi W_\mu^1 - \sin \varphi W_\mu^2, \nn
& \calW_\mu=\sin \varphi W_\mu^1+\cos \varphi W_\mu^2, \nn
& B_\mu=\cos \varphi' B_\mu^1 - \sin \varphi' B_\mu^2, \\
& \calB_\mu=\sin \varphi' B_\mu^1+\cos \varphi' B_\mu^2, \nonumber
\end{align}
with $t=\tan \varphi=g_1/g_2$ and $t'=\tan \varphi'=g_1^\prime/g_2^\prime$.
$W_\mu$ and $B_\mu$ are the massless (before electroweak breaking) gauge bosons of the SM gauge group, while the heavy vectors $\calW_\mu$  and $\calB_\mu$ have masses $\mu_\calW=\left| \frac{\Lambda g}{\cos \varphi \sin \varphi} \right|$ and $\mu_\calB=\left| \frac{\Lambda Y g'}{2\cos \varphi' \sin \varphi'} \right|$. Their couplings to fermions are $g_\calW^q=g_\calW^l=g t$, $g_\calB^\psi=g' t' Y^\psi$. Therefore, they are equal to the ones in the SM with a global rescaling by $t$ or $t^\prime$.

In model A, the couplings involving the Higgs doublet have the values $g_{\calW\calW}=g^2 t^2/4$, $g_{\calB\calB} = g'^2 t'^2/4$, $g_{\calW\calB}=g g' t t'$, $g_\calW^\phi=gt$, $g_\calB^\phi=g't'/2$. In model B, the formulas are obtained from this by the substitutions $t\to t^{-1}$, $t'\to t'^{-1}$. Therefore, in the configuration A the fermion and Higgs couplings are proportional, whereas in B they are inversely proportional. This difference has little impact on the direct searches below, so for brevity we only consider model A in the following.
We also assume that the couplings to the Higgs are not large and  $\mu_{\calW,\calB} \gg v$, such that we can neglect the mixing of the heavy vectors with the SM ones. 

Next, we make the crucial assumption $\mu_\calB=\mu_\calW$, which can be achieved for any value of $t$ and $t'$ by adjusting the hypercharge $Y$ of $\chi$. This tuning can be justified by some symmetry of the completion of our model. For instance, in models with extra dimensions it is a direct consequence of the higher-dimensional Lorentz invariance. For our purposes, it has the virtue of allowing for big mixing (regime 2). With this assumption, the mixing angle of $\calB_\mu$ with $\calW^3_\mu$ is given  by 
\beq
\tan \theta = \frac{g' t'}{g t} = \tan \theta_W \frac{t'}{t} .
\eeq
Notice that with this definition $\theta \in [0,\pi/2]$, unlike our choice for the general case discussed 
in Section \ref{sec:tripletsinglet}. This is convenient in this model as it allows for a well-defined identification of the two neutral
states with one Z-like and one $\gamma$-like extra vectors, which we call \Zp{} and $\gamma^\prime$. Only for $\theta \in [0,\pi/4]$ these states are equal, respectively, to the Z$^\prime_1$ and Z$^\prime_2$ of Section~\ref{sec:tripletsinglet}. The eigenmasses read%
\begin{align}
&M_{W'}^2=M_{\calW}^2=\mu_\calW^2 + \frac{g^2 t^2}{4} v^2, \label{massesGSM1} \\
&M_{Z'}^2= \mu_\calW^2 + \frac{g^2 t^2}{4\cos^2 \theta} v^2, \label{massesGSM2} \\
&M_{\gamma'}^2= \mu_\calW^2, \label{massesGSM3}
\end{align}
where we can observe the similarity with the pattern of masses of the SM gauge bosons. Moreover, the fermionic couplings of the heavy mass eigenstates W$^\prime$, Z$^\prime$, and $\gamma^\prime$ are equal to the ones of the W, Z and $\gamma$ bosons, respectively, with the replacements $g\to g t$, $\theta_W \to \theta$, with $e$ and $g'$ changing accordingly. Instead of using $t$,  it turns out to be more convenient to work in terms of $\tbar=(t+t^\prime)/2$. The parameterizations $(t,t^\prime)$ and $(\tbar,\theta)$ are then related as follows,
\beq
t=\frac{2~\!\tbar}{1+\frac{\tan{\theta}}{\tan{\theta_W}}},~~~~~t^\prime=\frac{2~\!\tbar}{1+\frac{\tan{\theta_W}}{\tan{\theta}}}.
\eeq
In particular, since $t$ and $t^\prime$ can always be chosen positive without loss of generality, this parameterization allows to easily keep track of the perturbativity of both the $\calW$ and $\calB$ couplings at the same time. At the particular point $\tbar=1$, $\theta=\theta_W$, this model has an exact replica of the SM gauge bosons, with a common shift of $\mu_\calW^2$ in their masses squared. Hence, in what follows we will refer to the model A with $\mu_\calB=\mu_\calW$ as the generalized sequential model (GSM).

\begin{figure}[t!]
\input{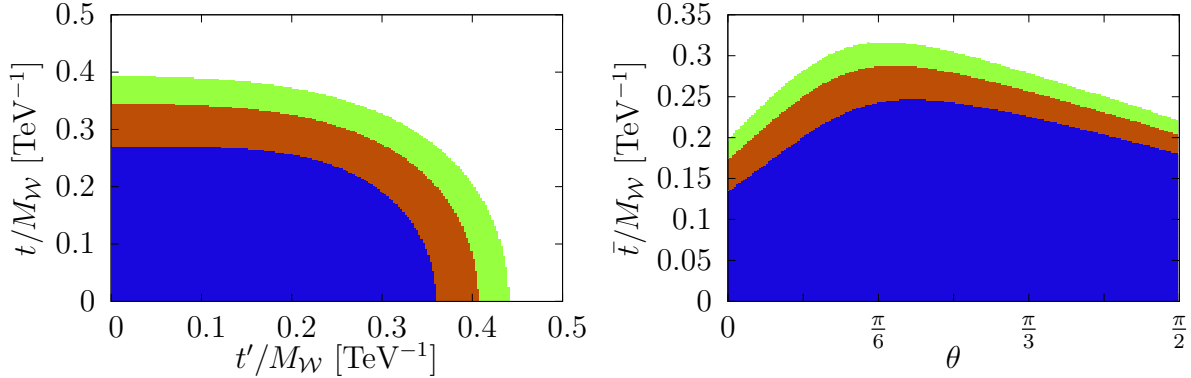}
\caption{Left: From darker to lighter, confidence regions with $\Delta \chi^2\leq2$ (blue), 4 (orange) and 6 ($95\%$ C.L.) (green), respectively, in the $t/M_{\calW}$-$t^\prime/M_{\calW}$ plane for the GSM. Right: The same using the parameterization in terms of $\tbar$ and $\theta$. 
}
\label{lim:GenSeq}
\end{figure}
As in the previous sections, we have calculated the bounds from electroweak precision data on this particular model. They are displayed in Figure \ref{lim:GenSeq}, using the two alternative parameterizations $(t,t^\prime)$ and $(\tbar,\theta)$.  We only find a small improvement in the minimum, $\chi^2-\chi^2_{\mathrm{SM}}=-0.48$. The individual $95\%$~C.L. limits on the individual parameters $t$ and $t^\prime$ are
\beq
\left|\frac{t}{M_{\calW}}\right|<0.34~\mathrm{TeV}^{-1}, ~~~~\left|\frac{t^\prime}{M_{\calW}}\right|<0.40~\mathrm{TeV}^{-1}.
\label{GSM_EWlim}
\eeq
Notice that there is no dependence on $\theta$ when $\tbar=0$, so this parameter cannot be bounded. Notice also that, within the dimension six effective Lagrangian expansion we are using to compute the limits, in both Figure \ref{lim:GenSeq} and Eq. (\ref{GSM_EWlim}) it is equivalent to express the results in terms of $t/M_\calW^2,~t^\prime/M_\calW^2$ or $t/\mu_\calW^2,~t^\prime/\mu_\calW^2$.

\begin{figure}[t!]
 \begin{center}
\input{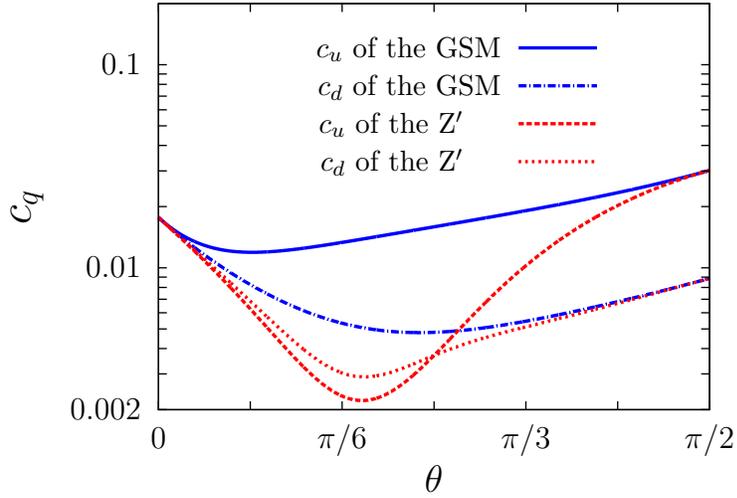}
 \caption{Values of the $c_u$ and $c_d$ parameters for the GSM with $\tbar=1$, as a function of $\theta$. Also shown are the $c_u$ and $c_d$ values for the (Z-like) Z$^\prime$ boson alone.}
 \label{CuCdGsm}
 \end{center}
\end{figure}
\begin{figure}[t!]
\begin{center}
\input{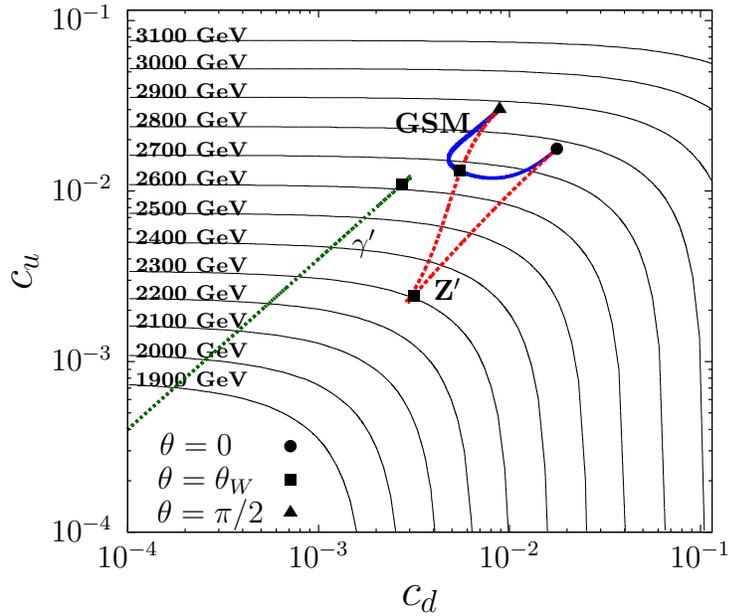}
\caption{$95\%$ C.L. exclusion limits for the cross section of a \Zp{}, for different masses, in the $c_u-c_d$ plane, taken from Ref.~\!\cite{Chatrchyan:2012it}. The $c_u$ and $c_d$ values of the GSM with $\tbar=1$ as a function of $\theta\in[0,\pi/2]$ are represented by the lines. We also plot the $c_u$-$c_d$ lines for the Z$^\prime$ and $\gamma^\prime$ bosons of the model, treated separately.}
\label{CuCdGsm_lim}
\end{center}
\end{figure}

We have used \refeq{cqint} to compute the effective $c_{u}-c_d$ parameters that describe the joint contribution of the two \Zp{} bosons to the dilepton cross section. The result is shown, for any $\theta$ and $\tbar=1$, in Figure \ref{CuCdGsm}. We also plot, for comparison, the usual $c_{u,d}$ values for Z$^\prime$, which when $\theta=\theta_W$ has the same couplings as the Z boson. 
In Figure \ref{CuCdGsm_lim}, we reproduce the CMS $95\%$ C.L. exclusion limits in the $c_u-c_d$ plane for different \Zp{} masses,  given in Ref.~\cite{Chatrchyan:2012it}. On top, we draw the line that represents the possible values of these parameters for a generalized sequential model with $\tbar=1$, together with the ones for the two \Zp{} treated independently. 
It is quite clear from these results that the predictions of a consistent sequential model are quite different from the ones of the benchmark that the experimental collaborations use for their interpretation of dilepton events. On the other hand, from Figure \ref{CuCdGsm_lim} we can read the following 95\%~C.L. limits on the masses when $\tbar=1$: $M_{Z^\prime} \leq 2800\GeV$, $2700\GeV$, $2900\GeV$ for $\theta=0$, $\theta_W$, $\pi/2$, respectively. As we will see shortly, these limits derived from the effective narrow-width approximation agree pretty well with the ones from a full calculation, which we describe in the next section. 

\section{Combined searches of \Zp{} and \Wp{} bosons at the LHC}
\label{sec:limits}

The LHC signals of singlets $\calB$ have been studied in great detail before. Here we present exclusion limits of extensions of the SM with at least one triplet $\calW$, using data from direct searches of resonances at the LHC. In these cases, we can have resonance signals with both $\ell^+ \ell^-$ and $\ell+\MET$ final states. To make use of the relations between \Zp{} and \Wp{} signals, we combine the data in both channels. This combination is carried out by defining a common test-statistic depending on the two channels. We will also compute, for comparison, the limits obtained from separate $\ell^+ \ell^-$ and $\ell+\MET$ events. We consider first extensions with one triplet only, and then put limits on the generalized sequential model, as an example of a model in Regime 2 (with large singlet-triplet mixing). Note that Regime 1 is covered by the triplet analysis and the standard searches of singlets.

We use the experimental data in the samples of $\ell^+ \ell^-$ and $\ell+\MET$ events collected by the CMS collaboration at $\sqrt{s}=7\TeV$ with an integrated luminosity $\mathrm{L_{int}}\sim5~\mathrm{fb}^{-1}$, as given in Refs.~\cite{Chatrchyan:2012it} and~\cite{Chatrchyan:2012meb}, respectively. The leptonic events include both electron and muon events. We generate the new physics signal using MadGraph/MadEvent 4.5 \cite{Alwall:2007st} together with PYTHIA 6.426 \cite{Sjostrand:2006za}.  Detector effects are simulated using PGS 4 \cite{pgs}. Our simulations also make use of a dedicated code to implement the matrix propagator of two nearby \Zp{} bosons, given in \refeq{propagator}. Finally, we take the background directly from Refs.~\cite{Chatrchyan:2012it} and~\cite{Chatrchyan:2012meb}. We have used in our simulations the same event selection cuts as the ones given in \cite{Chatrchyan:2012it} and \cite{Chatrchyan:2012meb}.

\begin{figure}[t!]
 \begin{center}
 \includegraphics[width=7cm]{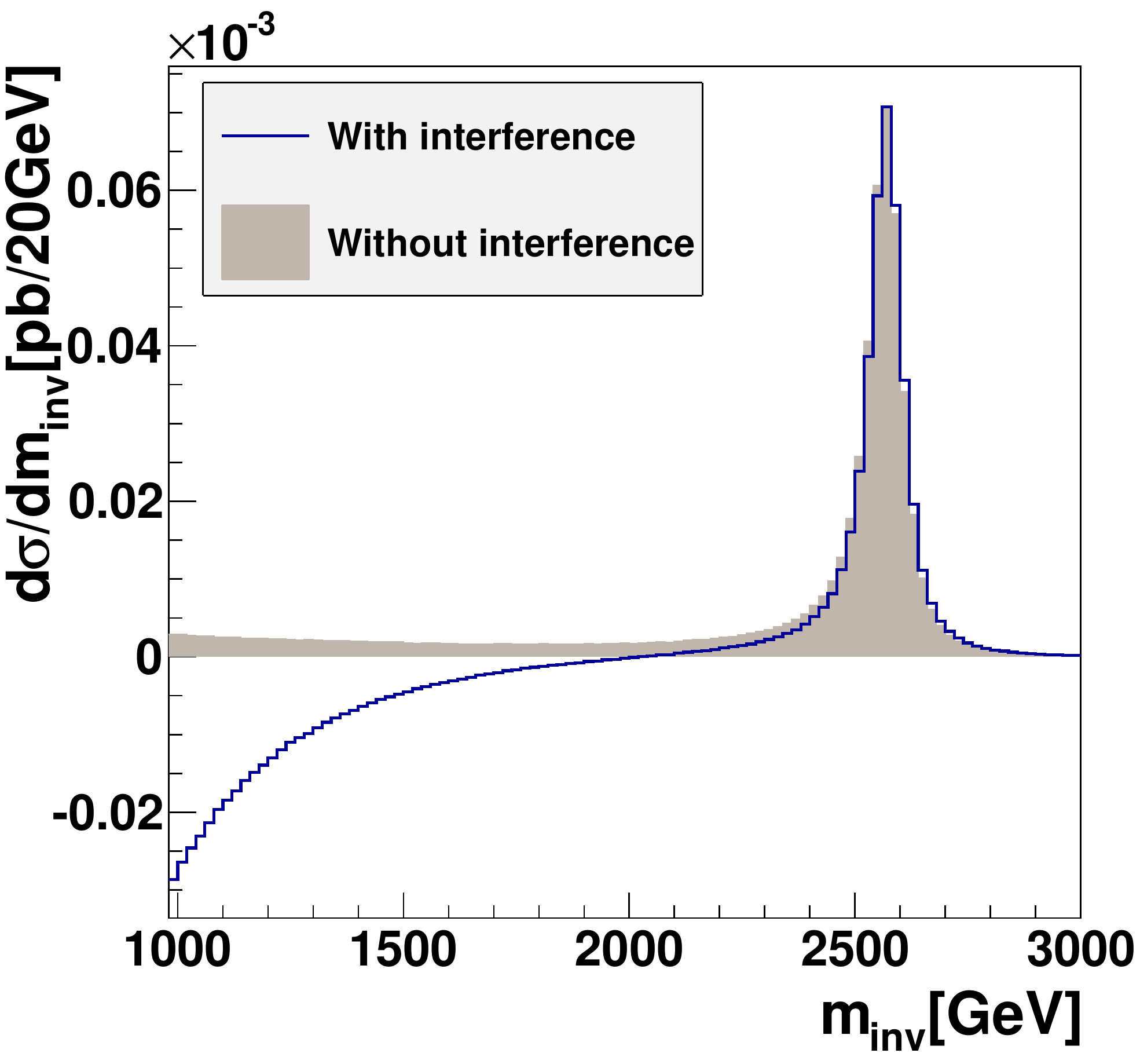}
 \includegraphics[width=7cm]{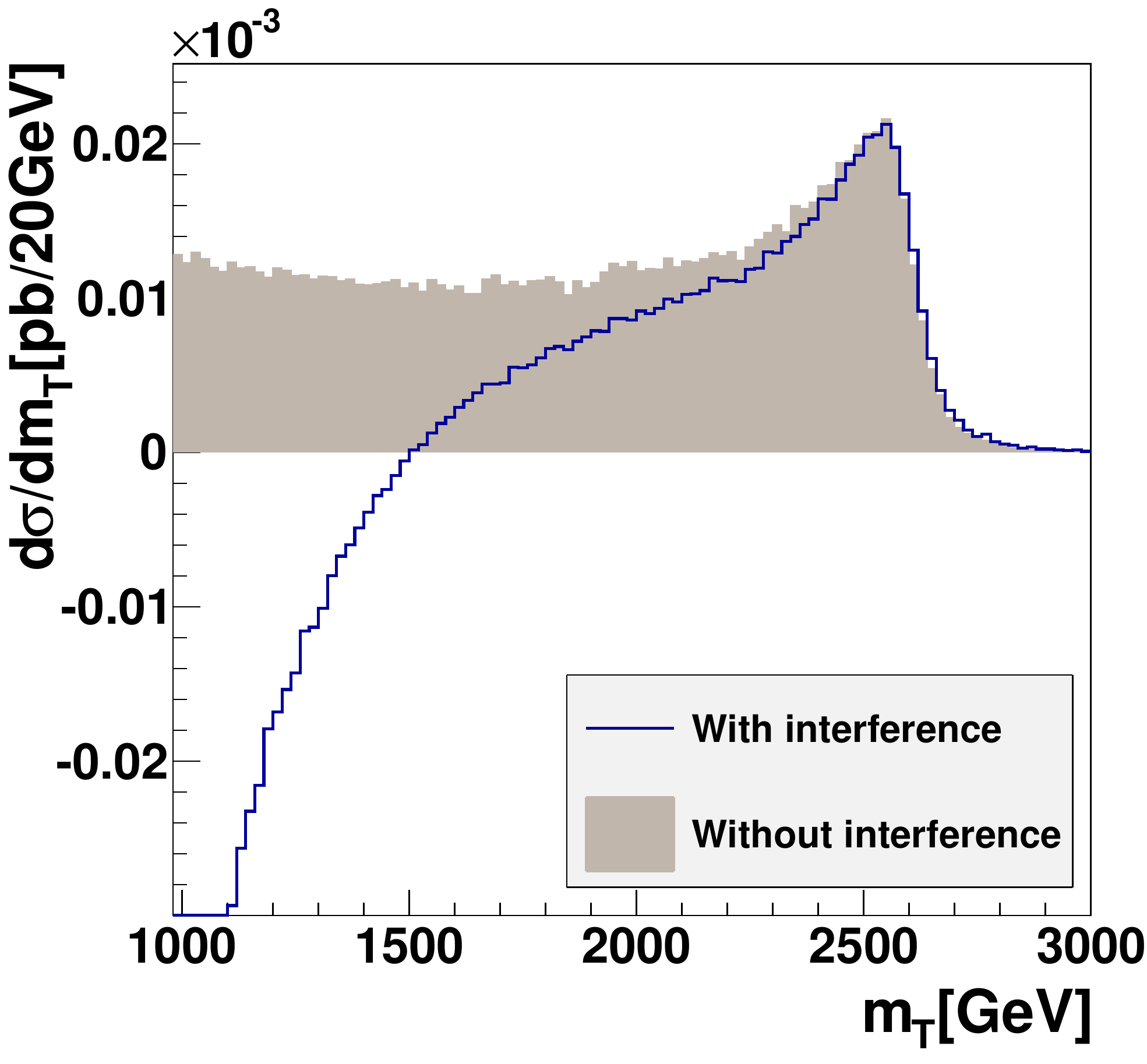}
 \caption{Comparison of the new physics contributions to the di-electron (left) and $e+\slashed{E}_T$ (right) channels, including and neglecting the interference with SM gauge bosons. The results correspond to a GSM with $M_{\calW}=2600\GeV$, $\theta=\theta_W$ and $\tbar=1$, for $\sqrt{s}=7\TeV$ at CMS. We use the cuts from \cite{Chatrchyan:2012it} and \cite{Chatrchyan:2012meb}.
 }
 \label{interference}
 \end{center}
\end{figure}

Since we neglect the mixing with SM gauge bosons, there is no production by gauge-boson fusion and we only need to consider the Drell-Yan processes $\bar{q}q \to \ell^+ \ell^-$ and $\bar{q}q \to \ell+\MET$. We observe that even in known models where the (colorless) extra vectors have rather small couplings to the light quarks and relatively large mixing, such as warped extra dimensions, Drell-Yan is the dominant production mechanism~\cite{Agashe:2007ki,Agashe:2008jb}. Our simulated signals allow for the possibility of emission of one jet. They also include the interference of the amplitudes involving new vector bosons with the ones mediated by SM gauge bosons. The importance of this effect is shown in Figure \ref{interference}, where we plot the excess of signal over background in the cross section distributions for a particular point of the generalized sequential model, in the $\ell^+ \ell^-$ and $\ell+\MET$ channels. The signal is calculated with and without the interference of the new amplitudes with the SM ones. It is apparent that the interference effects cannot be neglected in the regions away from the resonance, for $\ell^+ \ell^-$, or the Jacobian peak, for $\ell+\MET$.
The analyses are done using a window of masses larger than Min$\{0.7M,1900~\mathrm{GeV}\}$ for the neutral channels and Min$\{0.6M,1400~\mathrm{GeV}\}$ for the charged channels, with $M$ the mass of the \Zp{} or the \Wp{}. With these choices, the interference with the SM has a significant effect on the limits from the charged sector \cite{Accomando:2011eu}.

The statistical analysis is performed with the $CL_S$ method. For the test statistic we use the likelihood ratio
\begin{align}
Q &=\frac{L(s+b|\mathrm{obs})}{L(b|\mathrm{obs})} \nn
& =e^{-\sum_i (s)_i}\displaystyle\prod_i\left[ 1+\frac{(s)_i}{(b)_i} \right]^{n_i} ,
\end{align}
where $L(A|\mathrm{obs})$ are likelihood functions, $n_i$, $s_i$ and $b_i$ are, respectively, the observed number of events, the expected number of signal events and the expected number of background events, all in bin $i$, and the sum and product are carried out over the total number of bins in the invariant mass or transverse mass. In the combined analyses, we simply include the bins from $\ell^+ \ell^-$ events and $\ell+\MET$ events in the same $Q$.  Note that $Q$ is larger when the data are more consistent with the signal-plus-background hypothesis $s+b$, and smaller when they are more consistent with the background hypothesis $b$.
The confidence estimators are given by
\begin{align}
CL_{b}(Q_{\mathrm{obs}})=P_{b}(Q<Q_{\mathrm{obs}}),\\
CL_{s+b}(Q_{\mathrm{obs}})=P_{s+b}(Q<Q_{\mathrm{obs}}),\\
CL_s(Q_{\mathrm{obs}})=\frac{CL_{s+b}(Q_{\mathrm{obs}})}{CL_{b}(Q_{\mathrm{obs}})},
\end{align}
where $P_A(Q<Q_{\mathrm{obs}})$ is the probability, under the hypothesis $A$, of measuring in a pseudo-experiment a value of $Q$ smaller than the observed one, $Q_{\mathrm{obs}}$. The regions excluded at the 95\%~C.L. are defined by the points in parameter space for which $CL_s(Q_{\mathrm{obs}})\leq 0.05$.

In order to estimate the systematic theoretical error of our signal simulation, the backgrounds due to the tree-level processes $Z/\gamma^*\to \ell^+ \ell^-$ and $W^{\pm} \to \ell \nu$ have been simulated in the high invariant mass region. The physics of these processes is very similar to the one of \Zp{} and \Wp{}$^{\pm}$ processes, so the systematic errors can be expected to be roughly the same. We obtain a difference with the background prediction of \cite{Chatrchyan:2012it} and \cite{Chatrchyan:2012meb} of around $10\%-15\%$, which we take as our systematic error. We do not take into account possible correlations between the systematic errors in the two channels.

\subsection{Limits on triplet}
\label{sec:trana}

%
 \begin{figure}[t!]
  \begin{center}
 \input{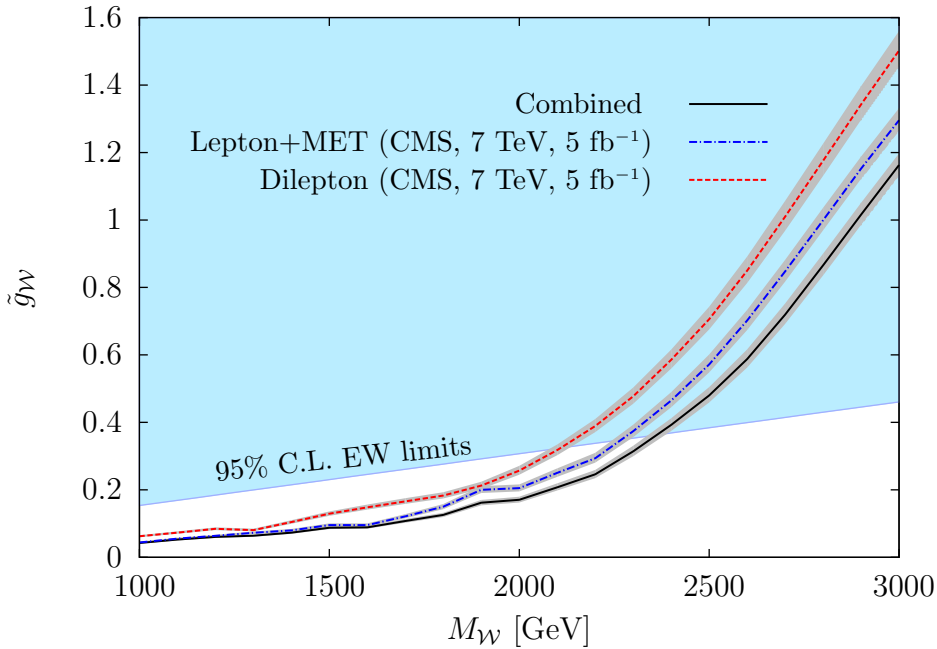}
  \caption{95$\%$ C.L. exclusion limits for the triplet model with $g^q_{\calW}=g^l_{\calW}=\tilde{g}_{\calW}$. Regions above the curves are excluded. The bounds obtained using the $\ell^+\ell^-$ and $\ell+\slashed{E}_T$ channels separately are delimited by the (red) dashed and (blue) dot-dashed lines, respectively. The solid black line represents the limits for the combination of both channels. Finally, the grey bands represent the systematic uncertainty, corresponding to a $\pm10\%$ variation in the signal. We also show in blue the region excluded by EWPD from a fit with $g_{\calW}^q=g_{\calW}^l$ and $g_{\calW}^\phi=0$. }
  \label{limitestr}
  \end{center}
 \end{figure}
As we have explained in Section~\ref{sec:triplet}, in the study of leptonic events at the LHC we can neglect the mixing of the new vectors with the Z and W bosons. Then, a general universal extension of the SM with a vector triplet $\calW$ is described by only three parameters (besides the ones in the SM): the mass $M_{\calW}^2$ and the couplings $g_{\calW}^q$ and $g_{\calW}^l$. We present our limits for the particular case $g_{\calW}^q=g_{\calW}^l \; (=\tilde{g}_\calW)$, where $\tilde{g}_\calW$ has been defined in~\refeq{gW}. In the narrow width approximation, the two couplings only enter the cross sections through the combination $\tilde{g}_\calW$, so these limits are also approximately valid for any value of $g_{\calW}^l/g_{\calW}^q$. 
 
Figure \ref{limitestr} shows our exclusion limits at the $95\%$ C.L. in the plane $M_{\calW}^2-\tilde{g}_{\calW}$ for the $\ell^+ \ell^-$ channel alone, the $\ell+\slashed{E}_T$ channel alone, and the combination of both channels, as explained above.
We can see that the $\ell+\slashed{E}_T$ data give stronger limits than the $\ell^+ \ell^-$ ones. The combined limits are thus dominated by the $\ell+\slashed{E}_T$ channel. Still, the combination gives a sizable improvement over this individual analysis. Let us emphasize that each of these curves put limits on the masses of both \WpL{} and \ZpL{} bosons. This means, for instance, that the information from $\ell+\slashed{E}_T$ alone already puts strong constraints on a \ZpL{} boson. These constraints are actually stronger than the ones from the more obvious dilepton channel, but weaker than the ones from the combination of the two channels. In Figure \ref{limitestr} we have also included the corresponding EWPD bounds, which still dominate the limits for large masses and couplings. In deriving the EWPD constraints we have assumed the same constraints, i.e. $g_{\calW}^q=g_{\calW}^l$ and no mixing with the SM ($g_{\calW}^\phi=0$). Notice that, while these bounds could be relaxed assuming free mixing (see Figure \ref{W0limits}), the effect of $g_{\calW}^\phi$ in the direct searches limits is small. 

\subsection{Limits on generalized sequential model}
\label{sec:gsmana}

A general extension of the SM model with one singlet and one triplet (neglecting mixing with the SM gauge bosons and supposing universality between families) depends on ten parameters. When $\Delta_{21} \gg v^2$ (Regime 1), the mixing angle $\theta$ approaches zero and it is possible to study the singlet and the triplet independently. The triplet analysis has been shown in the previous section, and the one for a singlet is well known. On the other hand, when $\Delta_{21} \lesssim v^2$ (Regime 2) the masses of all the extra particles are comparable, the mixing can be large, and we need to consider the triplet and the singlet at the same time.

We illustrate the special features of regime 2 with an analysis of the generalized sequential model, which has just three free parameters: $M_\calW$, $\tbar$ and $\theta$. In other words, we consider a three-dimensional slice of the general parameter space.
This model has two \Zp{} and one \Wp{}, all of them with a similar mass for any value of $\tbar$ and $\theta$. The interference and width-mixing effects between both \Zp{} bosons cannot be neglected in the calculation of the dilepton cross section. When $\theta=\theta_W$, the model contains a sequential \Wp, a sequential \Zp~and a sequential $\gamma^\prime$. 
\begin{figure}[t!]
 \begin{center}
\input{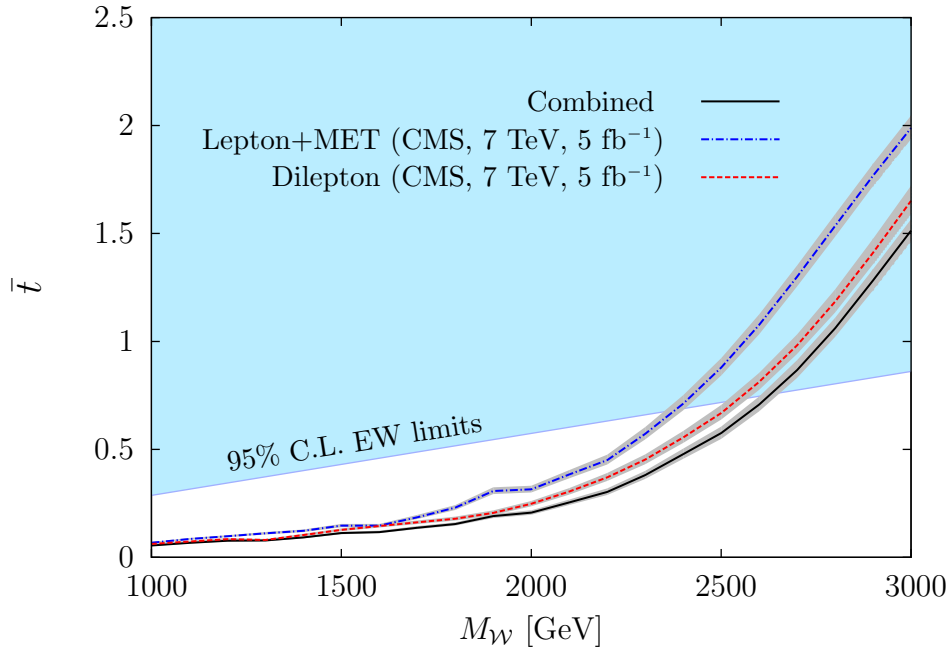}
 \caption{95$\%$ C.L. exclusion limits for the GSM with $\theta=\theta_W$. Regions above the curves are excluded. The dashed and dot-dashed lines delimit the bounds obtained using the $\ell^+\ell^-$ and $\ell+\slashed{E}_T$ channels separately, respectively. The solid line combines both of them. Systematic uncertainties, corresponding to a variation of a 10$\%$ of the signal, are represented by the grey bands. We also show in blue the region excluded by EWPD.}
 \label{limitest1}
 \end{center}
\end{figure}
\begin{figure}[t!]
 \begin{center}
\input{limtheta}
 \caption{95$\%$ C.L. exclusion limits  for the GSM with $\tbar=1$. Regions below the curves are excluded. The dashed and dot-dashed lines delimit the bounds obtained using the $\ell^+\ell^-$ and $\ell+\slashed{E}_T$ channels separately, respectively. The solid line combines both of them. Systematic uncertainties, corresponding to a variation of a 10$\%$ of the signal, are represented by the grey bands. We also show in blue the region excluded by EWPD at $95\%$ C.L. .
 }
 \label{limitesth}
 \end{center}
\end{figure}

In Figure \ref{limitest1} we show the exclusion limits at the $95\%$~C.L. in the plane $M_{\calW}-\tbar$ for the particular value $\theta=\theta_W$, from each separate data set and from their combination. From this plot, the limits on the individual masses of the two \Zp{} and the \Wp{} can be read directly, using Eqs.~(\ref{massesGSM1}-\ref{massesGSM3}). At the leading order, $M_{\mathrm{W^\prime}}\simeq M_{\mathrm{Z^\prime}}\simeq M_{\mathrm{\gamma^\prime}}\simeq M_{\calW}$. In Figure \ref{limitesth} we plot the limits in the plane $\theta-M_{\calW}$, for fixed $\tbar=1$. Again, we give the limits obtained from both channels separately and also from their combination. As in the triplet case, we also show the corresponding EWPD bounds in Figures \ref{limitest1} and  \ref{limitesth}, and these provide the stronger limits for large masses and couplings.

Comparing Figures \ref{CuCdGsm_lim} and \ref{limitesth}, we observe a good agreement between the limits found for the $\ell^+ \ell^-$ final state and the limits extracted from the curve of the $c_u-c_d$ plane. The allowed regions in that plane were obtained by CMS using the same set of data we employ. This shows that the generalization of the $c_u,~c_d$ parameters that we have proposed at the end of Section~\ref{sec:tripletsinglet} is accurate and useful.

For small values of $\theta$, the limits are dominated by the $\ell+\slashed{E}_T$ data, whereas when $\theta \gtrsim \pi/12$, the leptonic channel takes over. For large values of $\theta$ the triplet, and thus the \Wp{}, decouples. We see again that the combination improves the bounds in the region where the contribution of the triplet is non-negligible. Using the combined analysis, we find a lower bound on the mass of the sequential vector bosons \Wp{}, \Zp{} and $\gamma^\prime$ ($\theta=\theta_W$, $\tbar=1$) around $2750\GeV$.  It is also apparent that, for $\bar{t}=1$, the EWPD put stronger bounds than the LHC with 2011 data. The reason is that $\bar{t}=1$ gives an effective theory with pretty large fermionic couplings. In this case, non-trivial limits correspond to high masses, and the direct limits are penalized by the fast decrease of the parton distribution functions at large momentum fraction.

\section{Conclusions}
\label{sec:conclusions}

Consistency with the SM gauge invariance imposes strong constraints on the possible spectra and interactions of new particles in any new physics scenario. In particular, it demands that the extra particles furnish complete multiplets under $SU(3)_C \times SU(2)_L \times U(1)_Y$. This trivial observation has direct implications on how they will manifest in experiments. 
In this paper, we have studied the correlations that gauge invariance imposes between
processes with different leptonic final states in the searches of new vector bosons.  
We have built general effective Lagrangians that describe the propagation of the extra vectors and their interactions with the SM fields. These Lagrangians are gauge invariant and automatically incorporate these generic constraints. They provide a consistent model-independent description of this class of new physics in terms of the couplings and masses of the multiplets, which allows for a direct and natural interpretation of the experimental results.

Under mild assumptions, we have shown that any new resonant signal  $\ell+\MET$ arising from an extra charged vector boson \Wp{} must be accompanied
by a dilepton resonance from a companion \Zp{} in the multiplet\footnote{In Ref.~\cite{Babu:2011sd} similar observations have been derived from the requirement of perturbative unitarity. This is not surprising, since perturbative unitarity is tightly related to gauge invariance.}.  Moreover, the parameters controlling the strength of the interactions in both channels are related. Thus, when looking for extra charged vectors, it is natural to include both channels in a joint analysis that takes these correlations into account.  We have carried out such an analysis in different scenarios and, as expected, have obtained stronger limits than the ones from the independent analysis of each channel.

On a related note, it is worth emphasizing the importance of bounds from electroweak precision data in new physics studies. As we have seen, despite the strong limits that direct searches at the LHC impose on new physics, electroweak bounds are still competitive and actually dominate in regions of the parameter space with large couplings and masses.
For instance, the electroweak precision limits for a theoretically-consistent sequential \Zp{} model are stronger than the direct LHC limits with 2011 data.

On the other hand, our results can also be read from the point of view of the implications of a discovery. Should we find a signal of new physics in the $\ell+\MET$ channel, then we would expect a correlated signal in dilepton searches. In that case, in the same way we improve limits by combining both channels, the combination of channels should improve the significance of the discovery. Conversely, if a new neutral vectorial resonance is discovered, the details of its couplings and invariant-mass distributions can be related to the presence or absence of a resonance in the charged channel. Our gauge-invariant formalism would allow to extract the maximum of model-independent information from these observations.

The model-independent results for the parameters of the effective Lagrangian can be directly translated to specific models with extra vector bosons. This includes simple models such as the GSM studied here and the ones proposed in Refs.~\cite{Schmaltz:2010xr,Jezo:2012rm,Cao:2012ng}, and also more complete models such as little Higgs or extra-dimensional theories. For instance, the existence of correlations between \Zp{} and \Wp{} bosons, which we study here in detail, have been emphasized in Ref. \cite{Jezo:2012rm} for $SU(2)\times SU(2)\times U(1)$ models and in Ref.~\cite{Barducci:2012kk} in the context of a four dimensional composite Higgs model \cite{DeCurtis:2011yx}. Of course, explicit models usually bring about additional correlations, since they effectively correspond to a slice of the general parameter space. 

In this paper we have only considered the differential cross section of leptonic events, as a function of the invariant or  transverse mass. It would be interesting to analyze as well other observables, such as asymmetries and angular distributions. The correlations between the different observables can be studied with the very same effective Lagrangians we have written here, and should be taken into account to take full advantage of all the experimental data. To finish, let us point out that, even though the results of this paper focus on extra vectors and leptonic final states, analogous considerations apply in the case of resonant production of any other new particles, e.g. extra quarks \cite{delAguila:2000rc,AguilarSaavedra:2009es}, leptons \cite{delAguila:2008pw,AguilarSaavedra:2009ik} or scalars. 

\vspace{-0.25cm}
\section*{Acknowledments}
\vspace{-0.25cm}

It is a pleasure to thank J.A. Aguilar-Saavedra, M. Chala and J. Santiago for useful discussions. The work of J.B. has been supported in part by the U.S. National Science Foundation under Grant PHY-0905283-ARRA. The work of J.M.L. and M.P.V. has been supported by the MICINN project FPA2010-17915.


\newpage
\appendix

\section{The effective narrow width approximation for nearby resonances}
\label{app:beyonbreit}

Even if the fine structure of nearby resonances is interesting, often it will not be visible, due to the experimental resolution and/or limited statistics. In these cases, the experiments are only sensitive to the integrated cross section of the invariant mass distribution in the resonance region around $\bar{m}^2=\frac{1}{2} \mathrm{Tr}~\!M^2$.
 In the following, we show that if the widths are much smaller than the masses, we can approximate the total cross section due to the different particles with an effective narrow width approximation. For simplicity, we restrict ourselves to the case of two nearly degenerate neutral vector bosons.

We consider a Drell-Yan process $q\bar{q}\to Z'_i \to \ell^+\ell^-$, with the s-channel exchange of two \Zp{} bosons. Neglecting fermion masses, the parton-level cross section reads at the leading order,
\begin{equation}
\begin{gathered}
\sigma_{q\bar{q}\to \ell^+\ell^-}(p_{q},p_{\bar{q}},p_{l},p_{\bar{l}})=
\left\{ G_{ki}^q\Tr[\slashed{p}_{\bar{q}}\gamma^{\mu}\slashed{p}_q\gamma^{\sigma}]+ \hat{G}_{ki}^q\Tr[\slashed{p}_{\bar{q}}\gamma^{\mu}\slashed{p}_q\gamma^{\sigma}\gamma^5] \right\}\\
\times \left\{ G_{jm}^\ell\Tr[\slashed{p}_{\ell^-}\gamma^{\nu}\slashed{p}_{\ell^+}\gamma^{\rho}]+ \hat{G}_{jm}^\ell\Tr[\slashed{p}_{\ell^-}\gamma^{\nu}\slashed{p}_{\ell^+}\gamma^{\rho}\gamma^5] \right\}
P^{ij}_{\mu\nu}P^{\dag km}_{\sigma \rho},
\end{gathered}
\end{equation}
where $p_i$ is the momentum of the particle $i$,
\begin{equation}
\begin{gathered}
G^{f}_{ki}=(g_V^{f})_k (g_V^{f})_i+ (g_A^{f})_k (g_A^{f})_i=\frac{1}{2} \left[ (g_L^{f})_k (g_L^{f})_i+ (g_R^{f})_k (g_R^{f})_i \right],\\
\hat{G}^{f}_{ki}=(g_V^{f})_k (g_A^{f})_i+ (g_V^{f})_i (g_A^{f})_k=\frac{1}{2} \left[ (g_R^{f})_k (g_R^{f})_i - (g_L^{f})_k (g_L^{f})_i \right],
\label{def1}
\end{gathered}
\end{equation}
and
\begin{align}
& P_{ij}^{\mu\nu}=-i g^{\mu\nu} \, \Delta_{ij}(p^2),\\
& \Delta_{ij}(p^2)=\left[ p^2\delta_{ij}-M^2_{ij}-i\hat{\Sigma}_{ij}(p^2) \right]^{-1},
\label{def2}
\end{align}
with $p=p_{q}+p_{\bar{q}}=p_{\ell^-}+p_{\ell^+}$, and $\hat{\Sigma}_{ij} (p^2)=\mathrm{Im}~\!\Pi_{T,ij}(p^2)$. The leading contributions to $\hat{\Sigma}$ come from loops of the SM fermions. In the approximation of massless fermions,
\begin{equation}
\hat{\Sigma}_{ij} (p^2)=\frac{p^2}{12\pi}\suma_{f} G^{f}_{ij} \equiv p^2 \Sigma_{ij}.
\label{sigma}
\end{equation}
The terms with $\hat{G}^{f}_{ki}$ cancel out when the cross section is integrated over angles. After convolution with the parton distributions functions, the total cross section reads
\begin{equation}
\sigma_{pp\to Z^\prime_{i} \to \ell^+\ell^-} = \int \mathrm{d} p^2  \, \suma_{q=u,d} \frac{dL_{q\bar{q}}}{dp^2} \frac{p^2}{36\pi } \mathrm{Tr}[G^\ell\Delta^{\dag}(p^2)G^q\Delta(p^2)],
\label{expr}
\end{equation}
where $\frac{dL_{q\bar{q}}}{dp^2}$ are the parton luminosities.

For not very broad resonances, the
cross section is sharply peaked at a small region around the masses of the two bosons. Therefore, in calculating the total cross section, we can assume that $p^2\frac{dL_{q\bar{q}}}{dp^2}$ is approximately constant in this region, and substitute it by its value at $\bar{m}^2$:
\beq
\sigma_{pp\to Z^\prime_{i} \to \ell^+\ell^-}  \simeq \suma_{q=u,d} \left. \frac{dL_{q\bar{q}}}{dp^2}\right|_{\bar{m}^2} \frac{\bar{m}^2}{36\pi } \int_{-\infty}^\infty dp^2~\! \mathrm{Tr}[G^\ell\Delta^{\dag}(p^2)G^q\Delta(p^2)].
\label{xsappr}
\eeq
We have also extended the integral in $p^2$ to the interval $(-\infty,\infty)$.

The trace in the integrand of (\ref{xsappr}) can be rewritten as 
\begin{equation}
\mathrm{Tr}[G^\ell\Delta^\dag G^q \Delta]=\frac{\mathrm{Tr}[G^\ell A^\dag G^q A]}{|D|^2},
\label{fra}
\end{equation}
with
\begin{align}
& A=(1+i\widetilde{\Sigma})p^2-\widetilde{M}^2, \\
& D=\det [(p^2-M^2)-i\Sigma p^2],
\label{det}
\end{align}
where for any $2\times2$ matrix $R$, $\widetilde{R}_{ij} \equiv\epsilon^m_j \epsilon^n_i R_{mn}$ is its adjugate matrix.

On the real line, both the numerator and the denominator are positive real polynomials, of degree two and four, respectively. Thus, we can write them in the following way:
\begin{align}
& \mathrm{Tr}[G^\ell A^\dag G^q A](p^2)=b_q(p^2-u_q)(p^2-u_q^*),\\
& DD^{\dag}(p^2)=a(p^2-\omega_1)(p^2-\omega_1^*)(p^2-\omega_2)(p^2-\omega_2^*),
\end{align}
with $u_q$, $\omega_i$ the roots of the polynomials, $a=1+\mathcal{O}\left(\Sigma^2\right)$ and $b_q=\mathrm{Tr}[G^\ell G^q] \left[1+\mathcal{O}\left(\Sigma^2\right)\right]$.

The poles of (\ref{fra}) will be then given by the eigenvalues $\omega_i$ of the matrix $(1-i\Sigma)^{-1}M^2$. When $M^2$ is nearly degenerate, we can write $M^2=\bar{m}^2 (1+\delta)$, where the matrix $\delta$ is a small perturbation from the degenerate limit. The limit we are interested in then corresponds to the case $\mathcal{O}(\delta)<\mathcal{O}(\Sigma)$, i.e. mass splitting smaller than widths. Diagonalizing $(1-i\Sigma)^{-1}$ and treating $\delta$ as a perturbation we find 
\begin{equation}
\omega_i=\frac{\bar{m}^2}{1+\Sigma_i^2}(1-i\Sigma_i)(1+\delta_{ii})+\mathcal{O}(\delta ^2),
\label{w}
\end{equation}
where $\Sigma_i$ are the eigenvalues of $\Sigma$ and $\delta_{ii}$, the expected values of $\delta$ in the eigenvector of $\Sigma_i$.

On the other hand, the numerator of (\ref{fra}) can be rewritten in the following way:
\begin{equation}
\mathrm{Tr}[G^\ell A^\dag G^q A]=[\Tr(G^\ell G^q)+\Tr(\widetilde{\Sigma}G^\ell\widetilde{\Sigma}G^q)]p^4-2\Tr[G^qG^\ell\widetilde{M}^2]p^2+\Tr(\widetilde{M}^2G^\ell\widetilde{M}^2G^q).
\end{equation}
The zeros of the numerator are then given by
\begin{equation}
u_q=\frac{1+\beta_q}{1+\alpha_q}\bar{m}^2[1 \pm i\sqrt{\alpha_q}]+\mathcal{O}(\delta^2),
\label{u}
\end{equation}
where 
\beq
\alpha_q=\frac{\Tr(\widetilde{\Sigma}G^\ell\widetilde{\Sigma}G^q)}{\Tr(G^\ell G^q)},~~~~~
\beta_q=\frac{\Tr(\widetilde{\delta}G^\ell G^q)}{\Tr(G^\ell G^q)}.
\eeq

Once we know the analytical structure of the trace, it is straightforward to compute the integral in~(\ref{xsappr}) with the residue theorem. At the leading order, we find
\beq
\int_{-\infty}^{\infty} ds \frac{\mathrm{Tr}[G^\ell A^\dag G^q A]}{|D|^2}=\frac{\pi}{\bar{m}^2}\frac{\Tr(G^qG^\ell)}{\Tr(\Sigma)}\left( 1+\frac{\alpha_q}{\det\Sigma} \right)\, \left[1+\mathcal{O}(\delta^2/\Sigma^2)+\mathcal{O}(\Sigma)\right]\,.
\eeq
Therefore, the total cross section is approximated by,
\beq
\sigma_{q\bar{q}\to Z'_i \to \ell^+ \ell^-} =  \frac{1}{36~\!\Tr~\! \Sigma } \suma_{q=u,d} \frac{dL_{q\bar{q}}}{d\bar{m}^2} \left[ \mathrm{Tr}\left(G^\ell G^q\right)+\frac{\mathrm{Tr}\left(G^\ell \widetilde{\Sigma} G^q \widetilde{\Sigma}\right)}{\det \Sigma} \right].
\eeq
Comparing with (\ref{cucdcs}), written in terms of the parton luminosities, we can extract  the values of $c_u$ and $c_d$ that describe the ensemble of neutral vector bosons. Just as the standard ones for single resonances, they depend only on the couplings of the two bosons of the model\footnote{In some models it may happen that $\mathrm{det}\,\Sigma=0$. This signals the decoupling of one of the two \Zp{} from the SM fermions. In this case we recover the usual expression for $c_q$. Indeed, it can be shown that $\mathrm{det}\,\Sigma \ll 1$ implies $\mathrm{Tr}\left(G^\ell \widetilde{\Sigma} G^q \widetilde{\Sigma}\right)\sim\left(\mathrm{det}\,\Sigma\right)^2$, so in (\ref{cqint}) the term $\frac{\mathrm{Tr}\left(G^\ell \widetilde{\Sigma} G^q \widetilde{\Sigma}\right)}{\det \Sigma}$ cancels, and we get Eq.~(\ref{cucd}) again.}:
\beq
c_q=  \frac{1}{6\pi~\!\Tr~\!\Sigma} \left[ \mathrm{Tr}\left(G^\ell G^q\right)+\frac{\mathrm{Tr}\left(G^\ell \widetilde{\Sigma} G^q \widetilde{\Sigma}\right)}{\det \Sigma} \right].
\eeq
Note that, at the leading order, there is no dependence on $\delta$, i.e.\ on the splitting between the masses of the two \Zp{}. Therefore, Eq. (\ref{cqint}) allows us to read the limits on models involving two nearly-degenerate \Zp{}s from the generic bounds on the $c_u-c_d$ planes that are provided by the experimental collaborations. The basic assumption is that the resonances are narrow, but broader than the mass splitting.



\end{document}